\documentclass[aps, prl,twocolumn,superscriptaddress]{revtex4-1}

\usepackage{dcolumn}
\usepackage{bm}
\usepackage[T1]{fontenc}
\usepackage{braket}
\usepackage{amsmath}
\usepackage{amssymb}
\usepackage{nccmath}
\usepackage{mathrsfs}

\usepackage{pgf}   
\usepackage{graphicx}
\usepackage{xcolor}
\usepackage{lipsum}
\usepackage{array}
\newcolumntype{P}[1]{>{\centering\arraybackslash}p{#1}}
\let\oldsim\sim 
\renewcommand{\sim}{{\oldsim}}

\graphicspath{ {./images/} }

\begin{document}
	\title{A Feshbach resonance in collisions between ultracold ground state molecules}


\author{Juliana J. Park}
\affiliation{Research Laboratory of Electronics, MIT-Harvard Center for Ultracold Atoms,
Department of Physics, Massachusetts Institute of Technology, Cambridge, Massachusetts 02139, USA}

\author{Yu-Kun Lu}
\affiliation{Research Laboratory of Electronics, MIT-Harvard Center for Ultracold Atoms,
Department of Physics, Massachusetts Institute of Technology, Cambridge, Massachusetts 02139, USA}

\author{Alan O. Jamison}
\affiliation{Institute for Quantum Computing and Department of Physics \& Astronomy,
University of Waterloo, Waterloo, Ontario N2L 3G1, Canada}

\author{Timur V. Tscherbul}
\affiliation{Department of Physics, University of Nevada, Reno, Nevada, 89557, USA}

\author{Wolfgang Ketterle}
\affiliation{Research Laboratory of Electronics, MIT-Harvard Center for Ultracold Atoms,
Department of Physics, Massachusetts Institute of Technology, Cambridge, Massachusetts 02139, USA}

	\begin{abstract}
	
	Collisional resonances are an important tool which has been used to modify interactions in ultracold gases, for realizing novel Hamiltonians in quantum simulations \cite{bloch2012quantum}, for creating molecules from atomic gases \cite{chin2010feshbach} and for controlling chemical reactions. So far, such resonances have been observed for atom-atom collisions, atom-molecule collisions \cite{yang2019observation, wang2021magnetic, son2022control, knoop2009observation, zenesini2014resonant} and collisions between Feshbach molecules which are very weakly bound \cite{chin2005observation, wang2019observation, ferlaino2010collisions}. Whether such resonances exist for ultracold ground state molecules has been debated due to the possibly high density of states and/or rapid decay of the resonant complex \cite{mayle2012statistical, mayle2013scattering, christianen2019quasiclassical, christianen2019photoinduced, liu2022bimolecular}. Here we report a very pronounced and narrow (25 mG) Feshbach resonance in collisions between two ground state NaLi molecules. This molecular Feshbach resonance has two special characteristics. First, the collisional loss rate is enhanced by more than two orders of magnitude above the background loss rate which is saturated at the $p$-wave universal value, due to strong chemical reactivity. Second, the resonance is located at a magnetic field where two open channels become nearly degenerate. This implies the intermediate complex predominantly decays to the second open channel.  We describe the resonant loss feature using a model with coupled modes which is analogous to a Fabry–P\'erot cavity. Our observations prove the existence of long-lived coherent intermediate complexes even in systems without reaction barriers and open up the possibility of coherent control of chemical reactions.
	\end{abstract}
\textbf{\maketitle}

\section{Introduction}

Collisional resonances profoundly change the properties of ultracold gases. Magnetically tunable Feshbach resonances have been used to modify interactions between ultracold atoms from strong to weak and attractive to repulsive, as well as to coherently convert atomic gases into molecular gases \cite{chin2010feshbach}. Collisional resonances have become an important tool not only for creating novel Hamiltonians in quantum simulations \cite{bloch2012quantum}, but also for probing and understanding interatomic potentials and interactions.

It has been a long-standing goal for the rapidly advancing field of ultracold molecules to harness the power of collisional resonances. Ultracold molecules provide opportunities to study quantum state controlled chemistry \cite{krems2008cold, balakrishnan2016perspective}, quantum simulation \cite{micheli2006toolbox, capogrosso2010quantum, blackmore2018ultracold}, and quantum information processing \cite{ni2018dipolar, herrera2014infrared, hughes2020robust, sawant2020ultracold}. Recent progress in producing molecules from ultracold atoms \cite{NaLiGround,ni2008high,winkler2007coherent,danzl2010ultracold,park2015ultracold,danzl2008quantum,PhysRevA.92.062714} or directly laser cooling molecules \cite{shuman2010laser,anderegg2018laser} has laid the groundwork for achieving atom-like control of ultracold molecules.

For molecular systems, collisional resonances can provide microscopic information about collision complexes, and they can be used to suppress or enhance chemical reactions. However, so far Feshbach resonances have been observed only in two systems of atom-molecule collisions (NaK $+$ K \cite{yang2019observation, wang2021magnetic}, NaLi $+$ Na \cite{son2022control}), and for collisions involving Feshbach molecules, which are vibrationally excited molecules very close to the dissociation continuum and where the resonances are close to atomic Feshbach resonances \cite{chin2005observation, knoop2009observation, zenesini2014resonant, wang2019observation, ferlaino2010collisions}. It has even been an open question whether collisional resonances can be observed at all for ultracold ground state molecules due to the possibly high density of states and/or rapid decay of resonant states \cite{mayle2012statistical, mayle2013scattering, christianen2019quasiclassical, christianen2019photoinduced, liu2022bimolecular}. 

Here we report the observation of a pronounced, isolated Feshbach resonance in collisions between fermionic NaLi molecules in their ro-vibronic ground state. The magnetically tunable resonance is extremely narrow ($\sim 25$ mG) and enhances the loss rate by more than two orders of magnitude, providing strong evidence for a stable, long-lived collision complex. The existence of long-lived complexes in a molecular system of high reactivity such as NaLi is unexpected and has strong implications for controlling ultracold chemistry via scattering resonances. The long-lived state revealed by our experiments is coherently excited, whereas so far, all other observations of collisional complexes in molecule-molecule collisions \cite{hu2019direct,liu2020photo, gregory2021loss, gregory2021molecule, ComplexesNaKNaRb, bause2021collisions} are compatible with an incoherent population.

The observed resonance is special in two regards: In simple models, resonantly enhanced losses are only possible if the background loss rate is much smaller than the so-called universal limit \cite{idziaszek2010universal}. A loss rate near the universal limit implies almost complete inelastic loss at short-range and should suppress any long-lived resonant state. However, we observe loss rates close to the universal limit outside the narrow resonance. Second, the NaLi $+$ NaLi Feshbach resonance is observed at a specific magnetic field where two open channels become degenerate. It is possibly a new type of Feshbach resonance with a mechanism different from the Feshbach resonances observed so far in ultracold atomic systems. This mechanism cannot be realized in ultracold collisions of alkali metal atoms since the required single-particle level degeneracies do not occur at practicable field strengths. In contrast, degeneracies between two-particle threshold energies are commonly found in molecule-molecule collisions and have been used to engineer shielding interactions in ultracold KRb~+~KRb and CaF~+~CaF collisions \cite{matsuda2020resonant, schindewolf2022evaporation, anderegg2021observation}. Our results suggest that the new type of degeneracy-induced magnetic resonance could be ubiquitous in ultracold molecular physics, offering a powerful new mechanism for tuning intermolecular interactions with external electromagnetic fields. We explain the observed behavior with simple models.

\section{Experimental Results}

\begin{figure}
    \centering
	\includegraphics[width = 83mm, keepaspectratio]{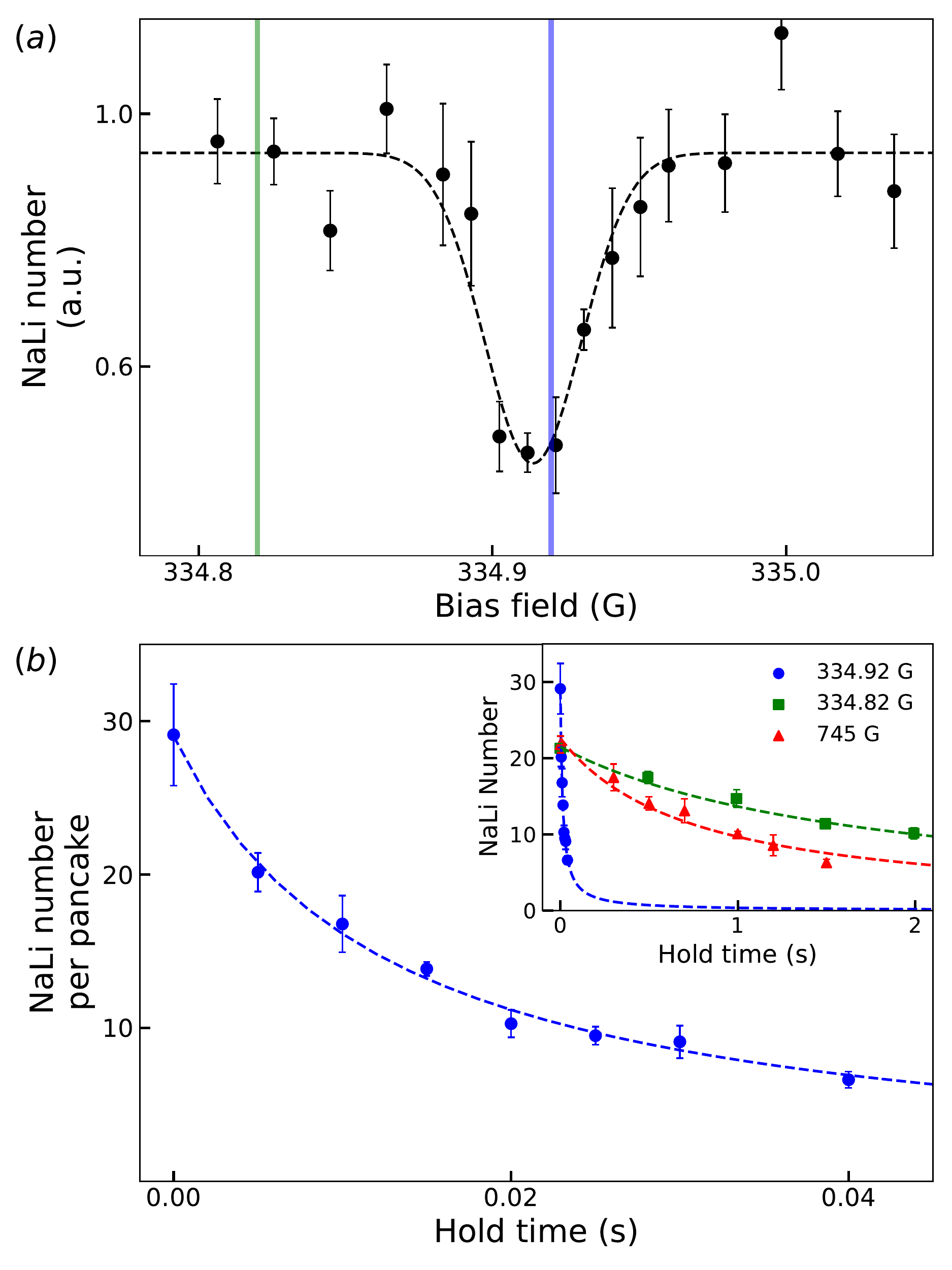}

	\caption{\textbf{a. Resonant molecular loss as a function of magnetic field.} Shown is the remaining molecule number after holding for $30 $ms at a target field near $334.9$ G. Molecule numbers are normalized by the molecule number without the $30$ ms hold. Each data point represents three to six measurements with and without $30$ ms hold respectively, and the error bars are one standard deviation. The black dashed line is a Gaussian fit. The blue (green) vertical line indicates the field where the molecular decay curve in blue (green) is obtained in Fig. \ref{fig:loss feature}(b).
	\textbf{b. Molecular decay curves at 334.92 G resonance and away from the resonance.} The main plot shows the decay curve near the center of the resonance within $50$ ms. The upper right subplot shows decay curves away from the resonance at 334.82 G in green squares and at 745 G in red triangles and also near the resonance at 334.92 G with blue circles from 0 to 2 s.
	Dashed lines are fits to a simple model for two-body loss using mean square regression (see Methods for the model for molecular decay).
	}
	\label{fig:loss feature}
\end{figure}

The experiment is done with $^{23}\text{Na}^{6}\text{Li}$ $ (a^{3}\Sigma ^{+})$ molecules in the ro-vibrational triplet ground state. The molecules are prepared with all spins aligned in the lower stretched hyperfine state. Following the techniques described in \cite{son2022control}, we prepare $6\times 10^4$ molecules at $1.8\;{\rm \mu K}$ temperature in a 1596 nm 1D optical lattice potential (further details in Methods). We search for scattering resonances in the bias field range $40.5\;{\rm G} <B< 1401.6\;{\rm G}$ while molecules are trapped in the 1D optical lattice. In this range of over $1300\;{\rm G}$, we observe a single Feshbach resonance with width $25\;{\rm mG}$ centered at $334.92\;{\rm G}$, as shown in Figure \ref{fig:loss feature}(a). The molecules become almost completely depleted at this field in 50 ms whereas more that half of the molecules survive at the background after 1 s, as shown is Figure \ref{fig:loss feature}(b). This corresponds to an increase in the NaLi loss rate by more than a factor of 100. We show in the following that the losses are due to two-body $p$-wave collisions, as expected for identical fermions. We investigate three aspects of the molecular decay rates: effect of impurities, density dependence, and temperature dependence.

\indent A small amount of impurities due to imperfect state preparation could cause rapid initial decay due to $s$-wave collisions that aren't suppressed by Pauli exclusion. However, after a rapid loss of these impurities, the fast decay would stop. By observing almost full decay of the molecular sample, we rule out that the fast decay is related to impurities. Figure \ref{fig:loss feature}(b) shows that about $30$ molecules per lattice site decay down to about 7 molecules in $40$ ms and are depleted to a barely detectable level in less than a hundred milliseconds. This confirms that the enhanced loss is due to collisions between fermionic molecules in a single state.

\begin{figure}
    \centering
	\includegraphics[width = 83mm, keepaspectratio]{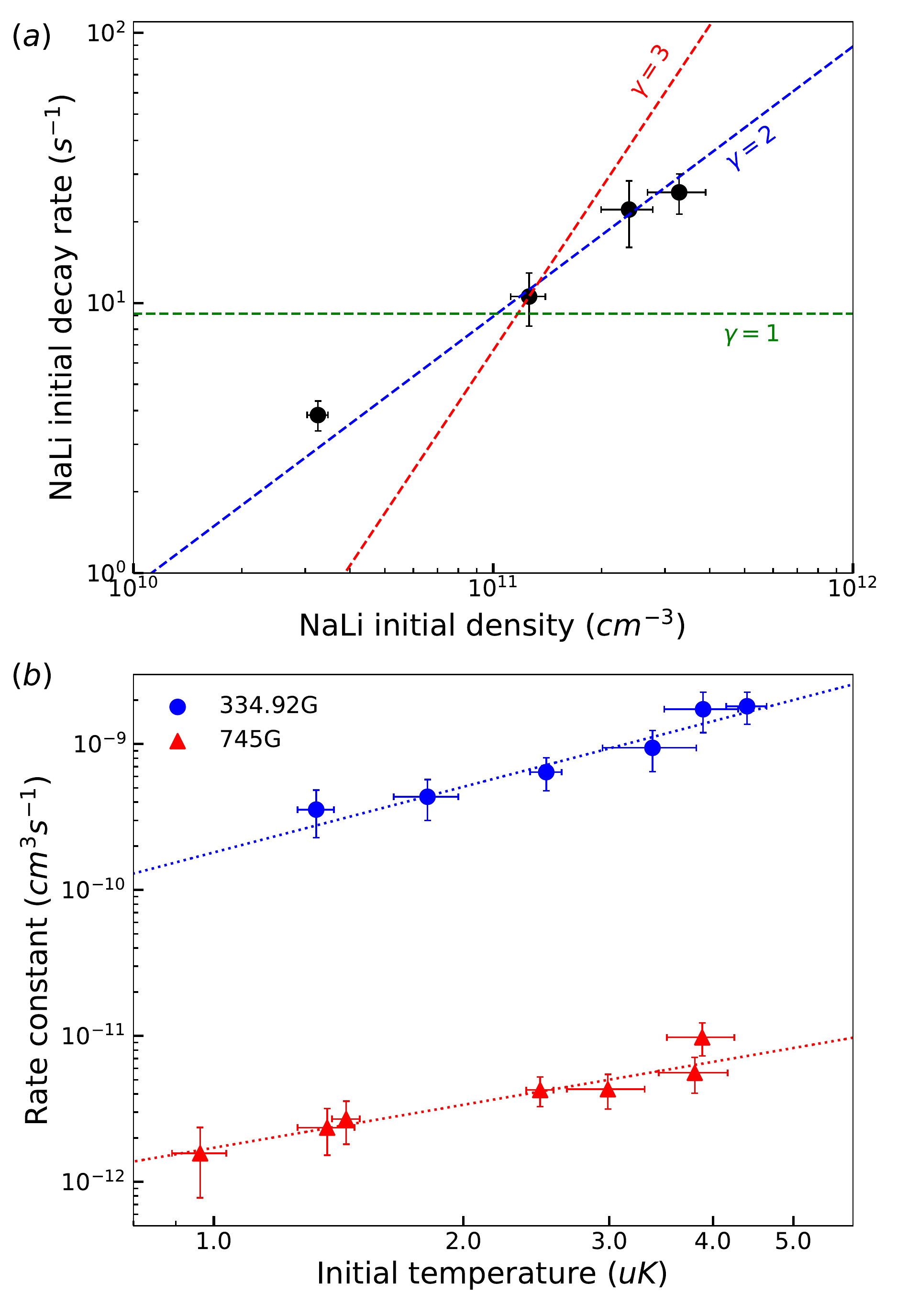}

	\caption{
	\textbf{a. Density dependency of molecular decay rate at $334.92$ G.} The initial decay rates are plotted as a function of initial molecule mean density. The green, blue and red dashed lines show the behavior expected for single molecule decay, two-body and three body collisions.
	\textbf{b. Threshold laws of molecule-molecule collisions.} Initial rate coefficients are plotted as a function of the initial temperature of NaLi molecules. Blue data points are measurements near the center of the resonance and red data points are measurements away from resonance near $745$ G. The lines show the linear dependence expected for $p$-wave collisions.}
	\label{fig:density and temperature}
\end{figure}

\indent To characterize the loss mechanism, the molecular decay rate $R_{o}$ is measured as a function of initial density $n_{o}$ and compared to the behavior $\beta n_{o}^{(\gamma-1)}$ expected for decay by collisions involving $\gamma$ particles. In general, the loss rate constant $\beta$ is temperature dependent. To avoid a more complicated analysis, we controlled the initial temperatures of the molecules to be the same within $15\%$.
Figure \ref{fig:density and temperature}(a), shows that the observed decay is due to binary collisions (a power-law fit gives $\gamma=1.85(9)$).
Molecule densities $n_{o}$ are estimated from the lattice trap frequencies and molecular temperature (see supplemental text for detail). 

\indent Next, we map out the temperature dependence of the molecular decay rate constant and compare with the Wigner threshold law \cite{wigner1948behavior}.
We generate molecular gases at different temperatures by varying the initial temperature of the Na and Li atomic mixture. Initial molecule temperatures ranging from $0.74(8)\;{\rm \mu K}$ to $3.88(36)\;{\rm \mu K}$ and from $1.33(7)\;{\rm \mu K}$ to $4.40(25)\;{\rm \mu K}$ are achieved away from ($745\;{\rm G}$) and at the resonance ($334.92\;{\rm G}$), respectively. The observed decay rate in Figure \ref{fig:density and temperature}(b) depends linearly on temperature as expected from the $p$-wave Wigner threshold law \cite{wigner1948behavior} for collisions between two identical fermions (a power-law fit $\beta = CT^{l}$ where $T$ is the temperature results in $l = 1.4(2)$ at the resonance, and $l = 0.98(19)$ at 745 G).

\indent

Natural comparisons for the observed decay rates are the unitarity limit and the universal loss rate. Our experiments in a 1D optical lattice are done in the crossover between 2D and 3D physics. For a quasi-2D trap, the unitarity limit is given by $\beta^{unit.}_{2D} {=} 4\frac{\hbar}{\mu}(\sqrt{\pi}l_{0})$ where $\mu$ is the reduced mass which is half of the NaLi molecule mass, $\mu=m_{\text{NaLi}}/2$ \cite{idziaszek2015reactive}. We see $\beta^{unit.}_{2D}$ scales linearly with the oscillator length in the tightly confined direction, $l_{0} = (\hbar/m_{\text{NaLi}}\omega_{z})^{1/2}$.
In contrast, the 3D unitarity limit neglects the zero-point motion due to 2D confinement and is given by $\beta^{unit.}_{3D} {=} 6\frac{\hbar}{\mu} \lambda_{dB}$ where $\lambda_{dB}$ is the thermal de Broglie wavelength $\lambda_{dB}{=}\sqrt{2\pi\hbar^2/k_B \mu T}$. Our experiments were carried out in the regime where the zero-point energy is larger than the thermal energy, and rates should be limited by the 2D limit since the 3D unitarity limit is higher. 
\indent To estimate the universal rate, we use an approximate value of the NaLi-NaLi long-range dispersion coefficient ($C_{6} =5879$ a.u) obtained by summing all $C_{6}$ coefficients between the two constituent atoms \cite{derevianko2001high}. The universal loss rate constant for $p$-wave ($s$-wave) collisions is $\beta^{univ}_{l=1}/T=1.2 \times 10^{-12}\;{\rm cm^{3}/s \cdot \mu K}$ ($\beta^{univ}_{l=0}=1.85 \times 10^{-10}\; {\rm cm^{3}/s}$). 
The background loss rate constants (at $745\;{\rm G}$) were obtained for various molecule temperatures and estimated to be $1.7(5) \times 10^{-12}\;{\rm cm^3/s} $ from a linear fit as shown by the red dotted line in Figure \ref{fig:density and temperature}(b). This background loss rate constant matches the $p$-wave universal value within the uncertainty.

\begin{figure}
        \centering
		\includegraphics[width = 83mm, keepaspectratio]{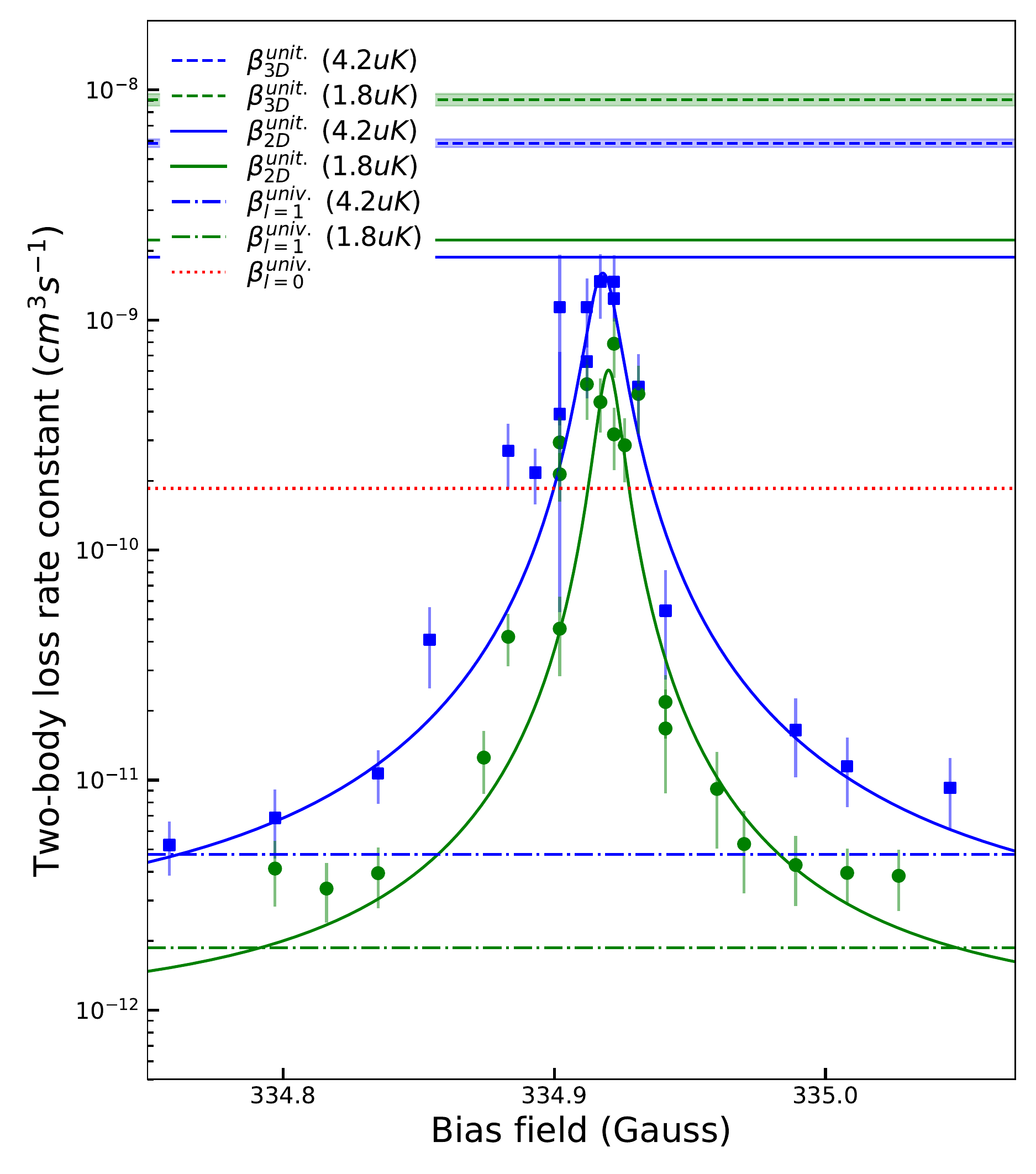}
		\caption{\textbf{Enhanced two-body loss rate coefficient of molecules.} Initial two-body loss rate coefficients which are indicated with blue squares and green circles are plotted as a function of bias field. Data points are obtained using a simple two-body loss model (see supplemental text) for initial temperatures $\sim 4.2$ uK (blue squares) and $\sim 1.8$ uK (green circles). Solid lines are Lorentzian fits to the data points. Dotted red line is the s-wave universal value and dash-dot lines are $p$-wave universal values for $T=4.2$ uK (blue) and $T=1.8$ uK (green). Dashed horizontal lines are the 3D unitarity limits and solid horizontal lines are the 2D unitarity limits.}
		\label{fig:loss feature combined}
    \end{figure}

The rate constant, $\beta$, increases by more than two orders of magnitude as the bias field approaches $334.918(5)\;\rm{G}$ from near the $p$-wave universal value to above the $s$-wave universal value. Loss rate coefficients are plotted as a function of magnetic field and fitted to a Lorentzian function for two temperatures, $1.8\;{\rm \mu K}$ and $4.2\;{\rm \mu K}$, in Fig. \ref{fig:loss feature combined}. The loss rate constant contrast is $\approx 150$ for $1.8\;{\rm \mu K}$ and $\approx 230$ for $4.2\;{\rm \mu K}$. The rate constants at the peaks are below the 3D unitarity limits but approach the 2D unitary limits.

The width of the resonance is comparable to the inhomogeneity of the magnetic field across the molecular sample, $\approx 25\;{\rm mG}$. Lorentian widths from the fits for both $1.8\;{\rm \mu K}$ and $4.2\;{\rm \mu K}$ are also $\approx 25\;{\rm mG}$. 
The result at $4.2\;{\rm \mu K}$ shows overall higher loss rate constant compared to that of $1.8\;{\rm \mu K}$, as expected from the $p$-wave threshold law at the resonance, $\beta \propto T$, mentioned earlier.

\begin{figure*}
    \centering
	\includegraphics[width = 170mm, keepaspectratio]{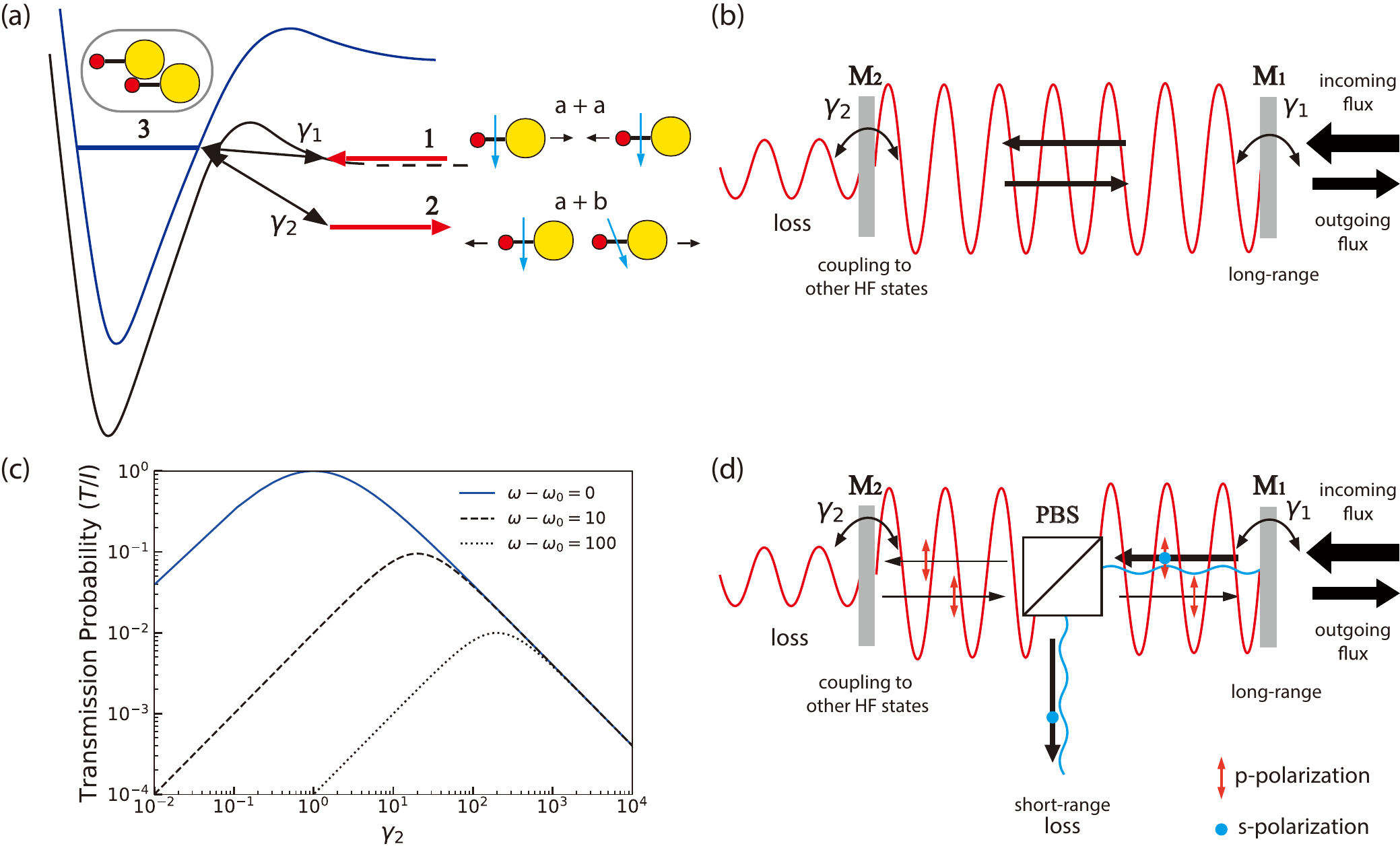}
	\caption{\textbf{a. Schematic of the resonance model with two open channels and a $p$-wave bound state trapped behind a centrifugal barrier.} Channel $\ket{1}$ and $\ket{2}$ are the open channels that are coupled to a closed channel $\ket{3}$, Channel $\ket{1}$ corresponds to the initial scattering channel where two NaLi molecules are in the lower stretched hyperfine state. Channel $\ket{2}$ corresponds to another open channel where molecules are in a different hyperfine state and energetically close to the incident scattering state. \textbf{b. Fabry–Pérot cavity model for molecular collisions.} Two mirrors M$_{1}$ and M$_{2}$ couple light into and out of the Fabry–Pérot resonator created by these mirrors with coupling strengths $\gamma_{i}$. When the spacing between the mirrors is tuned to form a cavity mode that is resonant to the incoming light, transmission loss is enhanced by constructive interference of light amplitudes inside the cavity. \textbf{c. Transmission probability of a cavity as a function of mirror coupling strengths.} For given coupling strengths of the first mirror, $\gamma_1=1$ (in arbitrary units), transmission probability reaches unity for resonant coupling (blue) at $\gamma_2$ $=$ $\gamma_1$ while maximizes to $\gamma_1/(\omega-\omega_0)$ for far off-resonant coupling (black) at $\gamma_2$ $=$ $2(\omega-\omega_0)$.
	\textbf{d. Schematic of a Fabry–Pérot cavity with a polarizing beam splitter (PBS).} Two mirrors form a resonator where the p-polarization component of the light amplitude is enhanced near a resonance. There are two paths for light to travel that are decoupled by the beam splitter. The s-polarization component is reflected by the PBS and lost from the cavity.}
	\label{fig:cartoon}
\end{figure*}

\section{Analysis}

\indent 
We will now develop a model which addresses our major experimental findings. The universal limit assumes that the loss rate is given by all the flux which has not been quantum reflected, i.e. it has tunneled through the centrifugal barrier located at $\approx 2.2 R_{\text{vdw}}$ with the van der Waals (vdW) length $R_{\text{vdw}}=\frac{1}{2}\left(\frac{2\mu C_{6}}{\hbar^2} \right)^{1/4} \approx 66 a_{0}$. Rates higher than the universal limit as observed here are only possible if the losses occur at long range (outside the $p$ wave barrier), or if substantial back reflection from behind the centrifugal barrier destructively interferes with the quantum reflection. Spin flip processes can in principle happen at long range due to the magnetic dipole-dipole interaction, but for the nearly degenerate hyperfine states in our system (see Methods for hyperfine structure of NaLi), the quantum numbers for the z-component of the electron spin differ by two from the input channel and therefore coupling by single spin flips is forbidden. Higher-order spin flips due to the intramolecular spin-spin and spin-rotation couplings are too weak: A loss rate constant higher than $10^{-10}\;{\rm cm^3 s^{-1}}$ requires a coupling strength on the order of $16\;{\rm kHz}$ at long range ($\sim 100\;{\rm nm}$). However, the strongest higher-order process has coupling strength around $0.05\;{\rm kHz}$ (further details in Methods). We therefore assume we have a $p$-wave resonance enabled by high reflectivity at close range.

In principle, the observed resonance could just be an ordinary $p$-wave resonance. However, the fact that this resonance occurs exactly at the magnetic field where the input channel becomes degenerate with another open channel suggests a new mechanism for which we will now introduce a minimum model. This model assumes two nearly degenerate states $\ket{1}$ and $\ket{2}$ coupled to a quasi-bound state $\ket{3}$ where two molecules are held together by the $p$-wave barrier and reflection at short range (see Fig. 4a). The rate of transfer of incoming flux $I$ in state $\ket{1}$ to state $\ket{2}$ depends on the coupling strengths $\gamma_{i}$ between channel $\ket{3}$ and the open channels $\ket{1}$ and $\ket{2}$ and on the energy difference between the incoming state $\hbar \omega$ and the quasi-bound state $\hbar \omega_{0}$: 

\begin{equation}   
    T_\text{trans}=I \cdot \frac{\gamma_{1}\gamma_{2}} {(\omega_o-\omega)^2+[(\gamma_{1}+\gamma_{2})/2]^2}.
    \label{eq:Lorenzian 1}
\end{equation} 
The process is fully analogous to an harmonic oscillator with resonance $\omega_0$ and damping rate $\gamma_2$ driven at frequency $\omega$ via a frictional input coupling $\gamma_1$. The power dissipated via $\gamma_2$ (normalized to quarter of the drive power at zero amplitude of the harmonic oscillator) is exactly given by Eq. \ref{eq:Lorenzian 1} (see Methods for details).

It is even more intuitive to generalize the optical analog which we have developed in a previous paper \cite{son2022control} which treated long-range quantum reflection and reflection at short range as the two mirrors of a Fabry-Perot interferometer. In our optical analog, the coupling $\gamma_2$ to the second open channel is represented by transmission through the inner mirror as shown in Figure \ref{fig:cartoon}(b). In the simplest model, we assume no additional loss for state 3 at short range.

Tuning the coupling term $\gamma_2$ now leads to a pronounced resonance-type feature in the Fabry-Perot transmission. For small $\gamma_2$, transmission is proportional to $\gamma_2$. However, for large $\gamma_2$, the quality factor of the resonance is reduced, there is less and less built up of light inside the resonator, and the transmission decreases with $\gamma_2$. For resonant coupling ($\omega-\omega_0 \ll \gamma_1$), the maximum transmission is $100 \%$ ($T_\text{trans}=I$) when $\gamma_2$ $=$ $\gamma_1$ while for off resonant coupling ($\omega-\omega_0 \gg \gamma_1$), the maximum transmission is at the reduced value of $\gamma_1/(\omega-\omega_0)$ when $\gamma_2$ $=$ $2(\omega-\omega_0)$ as shown in Figure \ref{fig:cartoon}(c).

We suggest that this mechanism is responsible for the observed resonance. For this, we assume that the incoming channel couples to a long-lived collision complex independent of the applied magnetic field. This would be the case for a $p$-wave shape resonance in the $F_{\rm tot}=7$ potential as the incident channel $\ket{aa,l=1,m_l=0}$ corresponds to the total angular momentum projection $M_\text{tot}=-7$. This may seem to call for a remarkable coincidence, but molecule-molecule collisions are predicted to have a high density of resonances \cite{mayle2012statistical,mayle2013scattering,christianen2019photoinduced,christianen2019quasiclassical}, so it is reasonable to assume that a near-threshold $p$-wave bound state is readily available at any magnetic field and collision energy.

We are using a generalization of the model used by \cite{chevy2005resonant} in the treatment of atomic $p$-wave Feshbach resonances which can be used to connect the parameters in Eq. (1) to a microscopic scattering theory (see Methods for details). The scaling of the decay strength $\gamma_2$ comes from the threshold behavior of $p$-wave inelastic collision rates ($\propto k'^{2}$) and the density of states of an open channel ($\propto k'$). When the external magnetic field is lowered, it tunes the channel $\ket{2}$ from above to below the energy of the input channel $\ket{1}$. Initially, channel $\ket{2}$ is closed and when it opens, it has zero-coupling strength due to the $k'^{3}$ term. For a magnetic field $\Delta B$ below the threshold, the coupling strength grows $\propto \Delta B^{3/2}$ which tunes the loss across its maximum as in Fig. 4(c).

Using a similar scaling for $\gamma_1 \propto k^3$, where $k$ is the wavevector in the input channel $\ket{1}$, we obtain the inelastic loss rate:

\begin{equation}    
   g_{2} \propto \frac{k^{-1}\gamma_1\gamma_2}{(E-E_0)^2+ [(\gamma_1+\gamma_2)/2]^{2}}
	\label{eq:Loss rate}
\end{equation} 
where $E$ is the incident energy and $E_0$ is the resonant energy. In the supplement, Eq. \ref{eq:Loss rate} is derived using a $T$-matrix formalism. Note that the numerator scales with $k^{2}$ instead of $k^{3}$, which is different in structure from Eq. \ref{eq:Lorenzian 1}. The reason is that for particle collisions, the initial channel has a specific wavevector, so we see threshold $p$-wave scaling ($k^2$) for the input coupling. Inelastic decay from the closed channel proceeds into a continuum of states, which adds another power of $k$ for the density of states in the continuum. The experimentally observed $k^2$ dependence of the loss rate constant implies that $\gamma_1$ never dominates the other terms in the denominator of $g_2$.

The model presented so far can explain a resonant enhancement of the loss as a function of magnetic field, but an enhancement by a factor of $1/y$ would lead to an off-resonant suppression of loss below the universal limit by a factor of $y$ based on the results of the quantum defect model presented in \cite{idziaszek2010universal, son2022control} (here $y$ is the standard short range absorption parameter). However, we observe a background loss comparable to the universal loss rate. One possible explanation is that the incoming flux is split into two or more (orthogonal) components, one part has the resonant behavior described above, whereas the other part has non-resonant universal loss (i.e. full transmission at short range). For example, if one tenth of the scattering flux is coupled to the resonance, but its loss is enhanced by a factor of 1000, the total loss can be 100 times the universal limit. Outside the resonance, the majority of the flux (here assumed to be 90 \%) provides the background loss near the universal rate. The optical analogy for this is the addition of a polarizing beam splitter to the Fabry-Perot interferometer, as shown in Figure \ref{fig:cartoon}(d). If most of the incoming flux matches the polarization that is reflected out of the cavity by the beam splitter, we would see a large background loss rate. Actually, for this polarization we have the optical analog of the universal limit where the ``inner mirror’’ is absent due to rapid short-range loss. If the molecule-molecule collision follows two or multiple paths split by some internal-state quantum numbers, this can explain the peculiarity of a large resonant enhancement above the universal limit from a background near to the universal limit.

For previously observed Feshbach resonances, one may estimate a lower limit for the lifetime of the collision complex by converting the observed width of the resonance to a lifetime. For our case of a $25\;\rm{mG}$ width we get $\sim 29\; \mu\rm{s}$, using a magnetic field sensitivity of 1 Bohr magneton to convert from magnetic field width to energy width. However, according to our model, the observed width may not reflect the width of the long-lived state but would depend instead on the ratio of coupling strengths for the two open channels. A more reliable estimate uses the classical round trip time $\tau_{RT}$ for a zero-energy particle in the combined centrifugal and vdW potential which is equal to $\tau_{RT}=\hbar/E_{\text{vdW}}\approx 5.6$ ns with the vdW energy, $E_{\text{vdW}}=\frac{\hbar^2}{2\mu}\frac{1}{R_{\text{vdw}}^2}$. The photon lifetime in a cavity is the round trip time times the resonant enhancement divided by 4. Using this analogy and the observed loss enhancement factor of 230, we obtain an estimate for the lifetime of the complex of 320 ns. 
We regard this as a low estimate. Even if the $p$-wave resonance is at resonance, due to thermal broadening, the maximum loss enhancement is reduced by a factor 
$\hbar \gamma_1/k_B T$. 
$\gamma_1$ is estimated by the classical round trip frequency $1/\tau_{RT}$ times the quantum transmission probability $\lvert t_1\rvert^2 \approx 2 \bar{a}_1 \bar{a}^2 k^3$ \cite{bai2019model}. For a collision energy of $4.2$ $\mu$K, this is approximately $2\pi \times 9$ kHz. Here, $\bar{a}=4 \pi R_\text{vdW}/\Gamma(\frac{1}{4})^2$, $\bar{a}_1=\bar{a}\Gamma(\frac{1}{4})^6$ and $k$ is the collision wavevector. 
Therefore, the reduction factor is approximately $0.1$, which implies a ten times larger lifetime. In addition, if we assume the loss enhancement is possibly not 230, but 2300 for $10\%$ of the incident flux (model with polarization beam splitter), then the lifetime would be another factor of ten higher and could be tens of microseconds. The observed enhancement of 230 provides an upper bound for the short-range absorption coefficient $y \leq 0.0043$.

The most surprising aspect of our results is the existence of a long-lived collision complex in NaLi$+$NaLi collisions, which allow barrier-free chemical reactions in all channels \cite{TripletReactions}. A long-lived complex is necessary for the existence of a high-Q resonance, such as we've observed, independently of any detailed model. Our observation of resonant behavior is very different from other recent experiments reporting long-lived collision complexes in molecule-molecule collisions, observed via photoionization \cite{hu2019direct} or inferred as the source of optical trap loss \cite{liu2020photo, gregory2021loss, gregory2021molecule, ComplexesNaKNaRb, bause2021collisions}. 
These observations were fitted to rate equations and are most likely related to ``sticky'' collisions \cite{mayle2012statistical, mayle2013scattering, christianen2019quasiclassical, christianen2019photoinduced, liu2022bimolecular} connected with a high density of states. Such complexes form incoherently and lead to a loss rate at the universal rate. In contrast, an enhancement above the universal rate (as observed here) is only possible when the flux reflected at short range destructively interferes with the quantum reflected flux and therefore requires full phase coherence.

In summary, we report an unexpected new type of Feshbach resonance in ultracold molecule-molecule collisions, the first observed between ground-state molecules at ultracold temperatures. The resonance rises up from a background loss at the universal limit, which is not possible with the most commonly studied models of ultracold collisions. It also implies the existence of a remarkably long-lived collision complex in a system with barrier-free reactions in all channels. The simplest such complex would be a $p$-wave shape resonance. As discussed above, we can put a lower bound on the lifetime of roughly $320\;{\rm ns}$, but it could be many times longer. The enhancement above a nearly universal loss implies the complex forms in a channel that may have small overlap with the incoming state. Additionally, the loss feature is almost certainly thermally broadened, further reducing the observed enhancement, and leading to a low estimate for the complex lifetime.

Although we explain the unusual features with a phenomenological model, they raise many questions for future work: Are narrow resonances such as ours unique to molecules with light atoms and therefore a lower density of states? Are resonant quasibound states coexisting with lossy channels a common feature of molecular systems which has so far gone undetected? In our case, we could detect the quasibound state only by modifying the coupling to a second, nearly degenerate open channel. Resonant states with low loss should have a distinct signature on elastic scattering properties which, however, may be difficult to measure in the presence of strong loss. Or does the non-observation of any other magnetically tunable molecule-molecule Feshbach resonance imply that those resonances are in general very dense, or broadened by strong coupling to other states or decay channels, and therefore not resolved? Our and other recent results \cite{liu2020photo,ComplexesNaKNaRb, gregory2021loss, gregory2021molecule, bause2021collisions} emphasize that the properties of collision complexes even for the simplest molecular systems are far from being understood.

\section{Acknowledgement}
We would like to thank John Bohn for valuable discussions. {\bf Funding:} We acknowledge support from the NSF through the Center for Ultracold Atoms and Grant No. 1506369 and from the Air Force Office of Scientific Research (MURI, Grant No. FA9550-21-1-0069). Some of the analysis was performed by W. K. at the Aspen Center for Physics, which is supported by National Science Foundation grant PHY-1607611. J. J. P. acknowledge additional support from the Samsung Scholarship. T.V.T. gratefully acknowledges support from the NSF CAREER award No. 2045681. {\bf Author contributions:} J. J. P carried out the experimental work. All authors contributed to the development of models, data analysis, and writing the manuscript. {\bf Competing interests:} The authors declare no competing interests. 

\clearpage
\bibliography{MolRes}

\begin{thebibliography}{60}%
\makeatletter
\providecommand \@ifxundefined [1]{%
 \@ifx{#1\undefined}
}%
\providecommand \@ifnum [1]{%
 \ifnum #1\expandafter \@firstoftwo
 \else \expandafter \@secondoftwo
 \fi
}%
\providecommand \@ifx [1]{%
 \ifx #1\expandafter \@firstoftwo
 \else \expandafter \@secondoftwo
 \fi
}%
\providecommand \natexlab [1]{#1}%
\providecommand \enquote  [1]{``#1''}%
\providecommand \bibnamefont  [1]{#1}%
\providecommand \bibfnamefont [1]{#1}%
\providecommand \citenamefont [1]{#1}%
\providecommand \href@noop [0]{\@secondoftwo}%
\providecommand \href [0]{\begingroup \@sanitize@url \@href}%
\providecommand \@href[1]{\@@startlink{#1}\@@href}%
\providecommand \@@href[1]{\endgroup#1\@@endlink}%
\providecommand \@sanitize@url [0]{\catcode `\\12\catcode `\$12\catcode
  `\&12\catcode `\#12\catcode `\^12\catcode `\_12\catcode `\%12\relax}%
\providecommand \@@startlink[1]{}%
\providecommand \@@endlink[0]{}%
\providecommand \url  [0]{\begingroup\@sanitize@url \@url }%
\providecommand \@url [1]{\endgroup\@href {#1}{\urlprefix }}%
\providecommand \urlprefix  [0]{URL }%
\providecommand \Eprint [0]{\href }%
\providecommand \doibase [0]{http://dx.doi.org/}%
\providecommand \selectlanguage [0]{\@gobble}%
\providecommand \bibinfo  [0]{\@secondoftwo}%
\providecommand \bibfield  [0]{\@secondoftwo}%
\providecommand \translation [1]{[#1]}%
\providecommand \BibitemOpen [0]{}%
\providecommand \bibitemStop [0]{}%
\providecommand \bibitemNoStop [0]{.\EOS\space}%
\providecommand \EOS [0]{\spacefactor3000\relax}%
\providecommand \BibitemShut  [1]{\csname bibitem#1\endcsname}%
\let\auto@bib@innerbib\@empty
\bibitem [{\citenamefont {Bloch}\ \emph {et~al.}(2012)\citenamefont {Bloch},
  \citenamefont {Dalibard},\ and\ \citenamefont
  {Nascimbene}}]{bloch2012quantum}%
  \BibitemOpen
  \bibfield  {author} {\bibinfo {author} {\bibfnamefont {I.}~\bibnamefont
  {Bloch}}, \bibinfo {author} {\bibfnamefont {J.}~\bibnamefont {Dalibard}}, \
  and\ \bibinfo {author} {\bibfnamefont {S.}~\bibnamefont {Nascimbene}},\
  }\href@noop {} {\bibfield  {journal} {\bibinfo  {journal} {Nature Physics}\
  }\textbf {\bibinfo {volume} {8}},\ \bibinfo {pages} {267} (\bibinfo {year}
  {2012})}\BibitemShut {NoStop}%
\bibitem [{\citenamefont {Chin}\ \emph {et~al.}(2010)\citenamefont {Chin},
  \citenamefont {Grimm}, \citenamefont {Julienne},\ and\ \citenamefont
  {Tiesinga}}]{chin2010feshbach}%
  \BibitemOpen
  \bibfield  {author} {\bibinfo {author} {\bibfnamefont {C.}~\bibnamefont
  {Chin}}, \bibinfo {author} {\bibfnamefont {R.}~\bibnamefont {Grimm}},
  \bibinfo {author} {\bibfnamefont {P.}~\bibnamefont {Julienne}}, \ and\
  \bibinfo {author} {\bibfnamefont {E.}~\bibnamefont {Tiesinga}},\ }\href@noop
  {} {\bibfield  {journal} {\bibinfo  {journal} {Reviews of Modern Physics}\
  }\textbf {\bibinfo {volume} {82}},\ \bibinfo {pages} {1225} (\bibinfo {year}
  {2010})}\BibitemShut {NoStop}%
\bibitem [{\citenamefont {Yang}\ \emph {et~al.}(2019)\citenamefont {Yang},
  \citenamefont {Zhang}, \citenamefont {Liu}, \citenamefont {Liu},
  \citenamefont {Nan}, \citenamefont {Zhao},\ and\ \citenamefont
  {Pan}}]{yang2019observation}%
  \BibitemOpen
  \bibfield  {author} {\bibinfo {author} {\bibfnamefont {H.}~\bibnamefont
  {Yang}}, \bibinfo {author} {\bibfnamefont {D.-C.}\ \bibnamefont {Zhang}},
  \bibinfo {author} {\bibfnamefont {L.}~\bibnamefont {Liu}}, \bibinfo {author}
  {\bibfnamefont {Y.-X.}\ \bibnamefont {Liu}}, \bibinfo {author} {\bibfnamefont
  {J.}~\bibnamefont {Nan}}, \bibinfo {author} {\bibfnamefont {B.}~\bibnamefont
  {Zhao}}, \ and\ \bibinfo {author} {\bibfnamefont {J.-W.}\ \bibnamefont
  {Pan}},\ }\href@noop {} {\bibfield  {journal} {\bibinfo  {journal} {Science}\
  }\textbf {\bibinfo {volume} {363}},\ \bibinfo {pages} {261} (\bibinfo {year}
  {2019})}\BibitemShut {NoStop}%
\bibitem [{\citenamefont {Wang}\ \emph {et~al.}(2021)\citenamefont {Wang},
  \citenamefont {Frye}, \citenamefont {Su}, \citenamefont {Cao}, \citenamefont
  {Liu}, \citenamefont {Zhang}, \citenamefont {Yang}, \citenamefont {Hutson},
  \citenamefont {Zhao}, \citenamefont {Bai} \emph {et~al.}}]{wang2021magnetic}%
  \BibitemOpen
  \bibfield  {author} {\bibinfo {author} {\bibfnamefont {X.-Y.}\ \bibnamefont
  {Wang}}, \bibinfo {author} {\bibfnamefont {M.~D.}\ \bibnamefont {Frye}},
  \bibinfo {author} {\bibfnamefont {Z.}~\bibnamefont {Su}}, \bibinfo {author}
  {\bibfnamefont {J.}~\bibnamefont {Cao}}, \bibinfo {author} {\bibfnamefont
  {L.}~\bibnamefont {Liu}}, \bibinfo {author} {\bibfnamefont {D.-C.}\
  \bibnamefont {Zhang}}, \bibinfo {author} {\bibfnamefont {H.}~\bibnamefont
  {Yang}}, \bibinfo {author} {\bibfnamefont {J.~M.}\ \bibnamefont {Hutson}},
  \bibinfo {author} {\bibfnamefont {B.}~\bibnamefont {Zhao}}, \bibinfo {author}
  {\bibfnamefont {C.-L.}\ \bibnamefont {Bai}},  \emph {et~al.},\ }\href@noop {}
  {\bibfield  {journal} {\bibinfo  {journal} {New Journal of Physics}\ }\textbf
  {\bibinfo {volume} {23}},\ \bibinfo {pages} {115010} (\bibinfo {year}
  {2021})}\BibitemShut {NoStop}%
\bibitem [{\citenamefont {Son}\ \emph {et~al.}(2022)\citenamefont {Son},
  \citenamefont {Park}, \citenamefont {Lu}, \citenamefont {Jamison},
  \citenamefont {Karman},\ and\ \citenamefont {Ketterle}}]{son2022control}%
  \BibitemOpen
  \bibfield  {author} {\bibinfo {author} {\bibfnamefont {H.}~\bibnamefont
  {Son}}, \bibinfo {author} {\bibfnamefont {J.~J.}\ \bibnamefont {Park}},
  \bibinfo {author} {\bibfnamefont {Y.-K.}\ \bibnamefont {Lu}}, \bibinfo
  {author} {\bibfnamefont {A.~O.}\ \bibnamefont {Jamison}}, \bibinfo {author}
  {\bibfnamefont {T.}~\bibnamefont {Karman}}, \ and\ \bibinfo {author}
  {\bibfnamefont {W.}~\bibnamefont {Ketterle}},\ }\href@noop {} {\bibfield
  {journal} {\bibinfo  {journal} {Science}\ }\textbf {\bibinfo {volume}
  {375}},\ \bibinfo {pages} {1006} (\bibinfo {year} {2022})}\BibitemShut
  {NoStop}%
\bibitem [{\citenamefont {Knoop}\ \emph {et~al.}(2009)\citenamefont {Knoop},
  \citenamefont {Ferlaino}, \citenamefont {Mark}, \citenamefont {Berninger},
  \citenamefont {Sch{\"o}bel}, \citenamefont {N{\"a}gerl},\ and\ \citenamefont
  {Grimm}}]{knoop2009observation}%
  \BibitemOpen
  \bibfield  {author} {\bibinfo {author} {\bibfnamefont {S.}~\bibnamefont
  {Knoop}}, \bibinfo {author} {\bibfnamefont {F.}~\bibnamefont {Ferlaino}},
  \bibinfo {author} {\bibfnamefont {M.}~\bibnamefont {Mark}}, \bibinfo {author}
  {\bibfnamefont {M.}~\bibnamefont {Berninger}}, \bibinfo {author}
  {\bibfnamefont {H.}~\bibnamefont {Sch{\"o}bel}}, \bibinfo {author}
  {\bibfnamefont {H.-C.}\ \bibnamefont {N{\"a}gerl}}, \ and\ \bibinfo {author}
  {\bibfnamefont {R.}~\bibnamefont {Grimm}},\ }\href@noop {} {\bibfield
  {journal} {\bibinfo  {journal} {Nature Physics}\ }\textbf {\bibinfo {volume}
  {5}},\ \bibinfo {pages} {227} (\bibinfo {year} {2009})}\BibitemShut {NoStop}%
\bibitem [{\citenamefont {Zenesini}\ \emph {et~al.}(2014)\citenamefont
  {Zenesini}, \citenamefont {Huang}, \citenamefont {Berninger}, \citenamefont
  {N{\"a}gerl}, \citenamefont {Ferlaino},\ and\ \citenamefont
  {Grimm}}]{zenesini2014resonant}%
  \BibitemOpen
  \bibfield  {author} {\bibinfo {author} {\bibfnamefont {A.}~\bibnamefont
  {Zenesini}}, \bibinfo {author} {\bibfnamefont {B.}~\bibnamefont {Huang}},
  \bibinfo {author} {\bibfnamefont {M.}~\bibnamefont {Berninger}}, \bibinfo
  {author} {\bibfnamefont {H.-C.}\ \bibnamefont {N{\"a}gerl}}, \bibinfo
  {author} {\bibfnamefont {F.}~\bibnamefont {Ferlaino}}, \ and\ \bibinfo
  {author} {\bibfnamefont {R.}~\bibnamefont {Grimm}},\ }\href@noop {}
  {\bibfield  {journal} {\bibinfo  {journal} {Physical Review A}\ }\textbf
  {\bibinfo {volume} {90}},\ \bibinfo {pages} {022704} (\bibinfo {year}
  {2014})}\BibitemShut {NoStop}%
\bibitem [{\citenamefont {Chin}\ \emph {et~al.}(2005)\citenamefont {Chin},
  \citenamefont {Kraemer}, \citenamefont {Mark}, \citenamefont {Herbig},
  \citenamefont {Waldburger}, \citenamefont {N{\"a}gerl},\ and\ \citenamefont
  {Grimm}}]{chin2005observation}%
  \BibitemOpen
  \bibfield  {author} {\bibinfo {author} {\bibfnamefont {C.}~\bibnamefont
  {Chin}}, \bibinfo {author} {\bibfnamefont {T.}~\bibnamefont {Kraemer}},
  \bibinfo {author} {\bibfnamefont {M.}~\bibnamefont {Mark}}, \bibinfo {author}
  {\bibfnamefont {J.}~\bibnamefont {Herbig}}, \bibinfo {author} {\bibfnamefont
  {P.}~\bibnamefont {Waldburger}}, \bibinfo {author} {\bibfnamefont {H.-C.}\
  \bibnamefont {N{\"a}gerl}}, \ and\ \bibinfo {author} {\bibfnamefont
  {R.}~\bibnamefont {Grimm}},\ }\href@noop {} {\bibfield  {journal} {\bibinfo
  {journal} {Physical review letters}\ }\textbf {\bibinfo {volume} {94}},\
  \bibinfo {pages} {123201} (\bibinfo {year} {2005})}\BibitemShut {NoStop}%
\bibitem [{\citenamefont {Wang}\ \emph {et~al.}(2019)\citenamefont {Wang},
  \citenamefont {Ye}, \citenamefont {Guo}, \citenamefont {Blume},\ and\
  \citenamefont {Wang}}]{wang2019observation}%
  \BibitemOpen
  \bibfield  {author} {\bibinfo {author} {\bibfnamefont {F.}~\bibnamefont
  {Wang}}, \bibinfo {author} {\bibfnamefont {X.}~\bibnamefont {Ye}}, \bibinfo
  {author} {\bibfnamefont {M.}~\bibnamefont {Guo}}, \bibinfo {author}
  {\bibfnamefont {D.}~\bibnamefont {Blume}}, \ and\ \bibinfo {author}
  {\bibfnamefont {D.}~\bibnamefont {Wang}},\ }\href@noop {} {\bibfield
  {journal} {\bibinfo  {journal} {Physical Review A}\ }\textbf {\bibinfo
  {volume} {100}},\ \bibinfo {pages} {042706} (\bibinfo {year}
  {2019})}\BibitemShut {NoStop}%
\bibitem [{\citenamefont {Ferlaino}\ \emph {et~al.}(2010)\citenamefont
  {Ferlaino}, \citenamefont {Knoop}, \citenamefont {Berninger}, \citenamefont
  {Mark}, \citenamefont {N{\"a}gerl},\ and\ \citenamefont
  {Grimm}}]{ferlaino2010collisions}%
  \BibitemOpen
  \bibfield  {author} {\bibinfo {author} {\bibfnamefont {F.}~\bibnamefont
  {Ferlaino}}, \bibinfo {author} {\bibfnamefont {S.}~\bibnamefont {Knoop}},
  \bibinfo {author} {\bibfnamefont {M.}~\bibnamefont {Berninger}}, \bibinfo
  {author} {\bibfnamefont {M.}~\bibnamefont {Mark}}, \bibinfo {author}
  {\bibfnamefont {H.-C.}\ \bibnamefont {N{\"a}gerl}}, \ and\ \bibinfo {author}
  {\bibfnamefont {R.}~\bibnamefont {Grimm}},\ }\href@noop {} {\bibfield
  {journal} {\bibinfo  {journal} {Laser Physics}\ }\textbf {\bibinfo {volume}
  {20}},\ \bibinfo {pages} {23} (\bibinfo {year} {2010})}\BibitemShut {NoStop}%
\bibitem [{\citenamefont {Mayle}\ \emph {et~al.}(2012)\citenamefont {Mayle},
  \citenamefont {Ruzic},\ and\ \citenamefont {Bohn}}]{mayle2012statistical}%
  \BibitemOpen
  \bibfield  {author} {\bibinfo {author} {\bibfnamefont {M.}~\bibnamefont
  {Mayle}}, \bibinfo {author} {\bibfnamefont {B.~P.}\ \bibnamefont {Ruzic}}, \
  and\ \bibinfo {author} {\bibfnamefont {J.~L.}\ \bibnamefont {Bohn}},\
  }\href@noop {} {\bibfield  {journal} {\bibinfo  {journal} {Physical Review
  A}\ }\textbf {\bibinfo {volume} {85}},\ \bibinfo {pages} {062712} (\bibinfo
  {year} {2012})}\BibitemShut {NoStop}%
\bibitem [{\citenamefont {Mayle}\ \emph {et~al.}(2013)\citenamefont {Mayle},
  \citenamefont {Qu{\'e}m{\'e}ner}, \citenamefont {Ruzic},\ and\ \citenamefont
  {Bohn}}]{mayle2013scattering}%
  \BibitemOpen
  \bibfield  {author} {\bibinfo {author} {\bibfnamefont {M.}~\bibnamefont
  {Mayle}}, \bibinfo {author} {\bibfnamefont {G.}~\bibnamefont
  {Qu{\'e}m{\'e}ner}}, \bibinfo {author} {\bibfnamefont {B.~P.}\ \bibnamefont
  {Ruzic}}, \ and\ \bibinfo {author} {\bibfnamefont {J.~L.}\ \bibnamefont
  {Bohn}},\ }\href@noop {} {\bibfield  {journal} {\bibinfo  {journal} {Physical
  Review A}\ }\textbf {\bibinfo {volume} {87}},\ \bibinfo {pages} {012709}
  (\bibinfo {year} {2013})}\BibitemShut {NoStop}%
\bibitem [{\citenamefont {Christianen}\ \emph
  {et~al.}(2019{\natexlab{a}})\citenamefont {Christianen}, \citenamefont
  {Karman},\ and\ \citenamefont {Groenenboom}}]{christianen2019quasiclassical}%
  \BibitemOpen
  \bibfield  {author} {\bibinfo {author} {\bibfnamefont {A.}~\bibnamefont
  {Christianen}}, \bibinfo {author} {\bibfnamefont {T.}~\bibnamefont {Karman}},
  \ and\ \bibinfo {author} {\bibfnamefont {G.~C.}\ \bibnamefont
  {Groenenboom}},\ }\href@noop {} {\bibfield  {journal} {\bibinfo  {journal}
  {Physical Review A}\ }\textbf {\bibinfo {volume} {100}},\ \bibinfo {pages}
  {032708} (\bibinfo {year} {2019}{\natexlab{a}})}\BibitemShut {NoStop}%
\bibitem [{\citenamefont {Christianen}\ \emph
  {et~al.}(2019{\natexlab{b}})\citenamefont {Christianen}, \citenamefont
  {Zwierlein}, \citenamefont {Groenenboom},\ and\ \citenamefont
  {Karman}}]{christianen2019photoinduced}%
  \BibitemOpen
  \bibfield  {author} {\bibinfo {author} {\bibfnamefont {A.}~\bibnamefont
  {Christianen}}, \bibinfo {author} {\bibfnamefont {M.~W.}\ \bibnamefont
  {Zwierlein}}, \bibinfo {author} {\bibfnamefont {G.~C.}\ \bibnamefont
  {Groenenboom}}, \ and\ \bibinfo {author} {\bibfnamefont {T.}~\bibnamefont
  {Karman}},\ }\href@noop {} {\bibfield  {journal} {\bibinfo  {journal}
  {Physical Review Letters}\ }\textbf {\bibinfo {volume} {123}},\ \bibinfo
  {pages} {123402} (\bibinfo {year} {2019}{\natexlab{b}})}\BibitemShut
  {NoStop}%
\bibitem [{\citenamefont {Liu}\ and\ \citenamefont
  {Ni}(2022)}]{liu2022bimolecular}%
  \BibitemOpen
  \bibfield  {author} {\bibinfo {author} {\bibfnamefont {Y.}~\bibnamefont
  {Liu}}\ and\ \bibinfo {author} {\bibfnamefont {K.-K.}\ \bibnamefont {Ni}},\
  }\href@noop {} {\bibfield  {journal} {\bibinfo  {journal} {Annual review of
  physical chemistry}\ }\textbf {\bibinfo {volume} {73}},\ \bibinfo {pages}
  {73} (\bibinfo {year} {2022})}\BibitemShut {NoStop}%
\bibitem [{\citenamefont {Krems}(2008)}]{krems2008cold}%
  \BibitemOpen
  \bibfield  {author} {\bibinfo {author} {\bibfnamefont {R.~V.}\ \bibnamefont
  {Krems}},\ }\href@noop {} {\bibfield  {journal} {\bibinfo  {journal}
  {Physical Chemistry Chemical Physics}\ }\textbf {\bibinfo {volume} {10}},\
  \bibinfo {pages} {4079} (\bibinfo {year} {2008})}\BibitemShut {NoStop}%
\bibitem [{\citenamefont {Balakrishnan}(2016)}]{balakrishnan2016perspective}%
  \BibitemOpen
  \bibfield  {author} {\bibinfo {author} {\bibfnamefont {N.}~\bibnamefont
  {Balakrishnan}},\ }\href@noop {} {\bibfield  {journal} {\bibinfo  {journal}
  {The Journal of chemical physics}\ }\textbf {\bibinfo {volume} {145}},\
  \bibinfo {pages} {150901} (\bibinfo {year} {2016})}\BibitemShut {NoStop}%
\bibitem [{\citenamefont {Micheli}\ \emph {et~al.}(2006)\citenamefont
  {Micheli}, \citenamefont {Brennen},\ and\ \citenamefont
  {Zoller}}]{micheli2006toolbox}%
  \BibitemOpen
  \bibfield  {author} {\bibinfo {author} {\bibfnamefont {A.}~\bibnamefont
  {Micheli}}, \bibinfo {author} {\bibfnamefont {G.}~\bibnamefont {Brennen}}, \
  and\ \bibinfo {author} {\bibfnamefont {P.}~\bibnamefont {Zoller}},\
  }\href@noop {} {\bibfield  {journal} {\bibinfo  {journal} {Nature Physics}\
  }\textbf {\bibinfo {volume} {2}},\ \bibinfo {pages} {341} (\bibinfo {year}
  {2006})}\BibitemShut {NoStop}%
\bibitem [{\citenamefont {Capogrosso-Sansone}\ \emph
  {et~al.}(2010)\citenamefont {Capogrosso-Sansone}, \citenamefont {Trefzger},
  \citenamefont {Lewenstein}, \citenamefont {Zoller},\ and\ \citenamefont
  {Pupillo}}]{capogrosso2010quantum}%
  \BibitemOpen
  \bibfield  {author} {\bibinfo {author} {\bibfnamefont {B.}~\bibnamefont
  {Capogrosso-Sansone}}, \bibinfo {author} {\bibfnamefont {C.}~\bibnamefont
  {Trefzger}}, \bibinfo {author} {\bibfnamefont {M.}~\bibnamefont
  {Lewenstein}}, \bibinfo {author} {\bibfnamefont {P.}~\bibnamefont {Zoller}},
  \ and\ \bibinfo {author} {\bibfnamefont {G.}~\bibnamefont {Pupillo}},\
  }\href@noop {} {\bibfield  {journal} {\bibinfo  {journal} {Physical review
  letters}\ }\textbf {\bibinfo {volume} {104}},\ \bibinfo {pages} {125301}
  (\bibinfo {year} {2010})}\BibitemShut {NoStop}%
\bibitem [{\citenamefont {Blackmore}\ \emph {et~al.}(2018)\citenamefont
  {Blackmore}, \citenamefont {Caldwell}, \citenamefont {Gregory}, \citenamefont
  {Bridge}, \citenamefont {Sawant}, \citenamefont {Aldegunde}, \citenamefont
  {Mur-Petit}, \citenamefont {Jaksch}, \citenamefont {Hutson}, \citenamefont
  {Sauer} \emph {et~al.}}]{blackmore2018ultracold}%
  \BibitemOpen
  \bibfield  {author} {\bibinfo {author} {\bibfnamefont {J.~A.}\ \bibnamefont
  {Blackmore}}, \bibinfo {author} {\bibfnamefont {L.}~\bibnamefont {Caldwell}},
  \bibinfo {author} {\bibfnamefont {P.~D.}\ \bibnamefont {Gregory}}, \bibinfo
  {author} {\bibfnamefont {E.~M.}\ \bibnamefont {Bridge}}, \bibinfo {author}
  {\bibfnamefont {R.}~\bibnamefont {Sawant}}, \bibinfo {author} {\bibfnamefont
  {J.}~\bibnamefont {Aldegunde}}, \bibinfo {author} {\bibfnamefont
  {J.}~\bibnamefont {Mur-Petit}}, \bibinfo {author} {\bibfnamefont
  {D.}~\bibnamefont {Jaksch}}, \bibinfo {author} {\bibfnamefont {J.~M.}\
  \bibnamefont {Hutson}}, \bibinfo {author} {\bibfnamefont {B.}~\bibnamefont
  {Sauer}},  \emph {et~al.},\ }\href@noop {} {\bibfield  {journal} {\bibinfo
  {journal} {Quantum Science and Technology}\ }\textbf {\bibinfo {volume}
  {4}},\ \bibinfo {pages} {014010} (\bibinfo {year} {2018})}\BibitemShut
  {NoStop}%
\bibitem [{\citenamefont {Ni}\ \emph {et~al.}(2018)\citenamefont {Ni},
  \citenamefont {Rosenband},\ and\ \citenamefont {Grimes}}]{ni2018dipolar}%
  \BibitemOpen
  \bibfield  {author} {\bibinfo {author} {\bibfnamefont {K.-K.}\ \bibnamefont
  {Ni}}, \bibinfo {author} {\bibfnamefont {T.}~\bibnamefont {Rosenband}}, \
  and\ \bibinfo {author} {\bibfnamefont {D.~D.}\ \bibnamefont {Grimes}},\
  }\href@noop {} {\bibfield  {journal} {\bibinfo  {journal} {Chemical science}\
  }\textbf {\bibinfo {volume} {9}},\ \bibinfo {pages} {6830} (\bibinfo {year}
  {2018})}\BibitemShut {NoStop}%
\bibitem [{\citenamefont {Herrera}\ \emph {et~al.}(2014)\citenamefont
  {Herrera}, \citenamefont {Cao}, \citenamefont {Kais},\ and\ \citenamefont
  {Whaley}}]{herrera2014infrared}%
  \BibitemOpen
  \bibfield  {author} {\bibinfo {author} {\bibfnamefont {F.}~\bibnamefont
  {Herrera}}, \bibinfo {author} {\bibfnamefont {Y.}~\bibnamefont {Cao}},
  \bibinfo {author} {\bibfnamefont {S.}~\bibnamefont {Kais}}, \ and\ \bibinfo
  {author} {\bibfnamefont {K.~B.}\ \bibnamefont {Whaley}},\ }\href@noop {}
  {\bibfield  {journal} {\bibinfo  {journal} {New Journal of Physics}\ }\textbf
  {\bibinfo {volume} {16}},\ \bibinfo {pages} {075001} (\bibinfo {year}
  {2014})}\BibitemShut {NoStop}%
\bibitem [{\citenamefont {Hughes}\ \emph {et~al.}(2020)\citenamefont {Hughes},
  \citenamefont {Frye}, \citenamefont {Sawant}, \citenamefont {Bhole},
  \citenamefont {Jones}, \citenamefont {Cornish}, \citenamefont {Tarbutt},
  \citenamefont {Hutson}, \citenamefont {Jaksch},\ and\ \citenamefont
  {Mur-Petit}}]{hughes2020robust}%
  \BibitemOpen
  \bibfield  {author} {\bibinfo {author} {\bibfnamefont {M.}~\bibnamefont
  {Hughes}}, \bibinfo {author} {\bibfnamefont {M.~D.}\ \bibnamefont {Frye}},
  \bibinfo {author} {\bibfnamefont {R.}~\bibnamefont {Sawant}}, \bibinfo
  {author} {\bibfnamefont {G.}~\bibnamefont {Bhole}}, \bibinfo {author}
  {\bibfnamefont {J.~A.}\ \bibnamefont {Jones}}, \bibinfo {author}
  {\bibfnamefont {S.~L.}\ \bibnamefont {Cornish}}, \bibinfo {author}
  {\bibfnamefont {M.}~\bibnamefont {Tarbutt}}, \bibinfo {author} {\bibfnamefont
  {J.~M.}\ \bibnamefont {Hutson}}, \bibinfo {author} {\bibfnamefont
  {D.}~\bibnamefont {Jaksch}}, \ and\ \bibinfo {author} {\bibfnamefont
  {J.}~\bibnamefont {Mur-Petit}},\ }\href@noop {} {\bibfield  {journal}
  {\bibinfo  {journal} {Physical Review A}\ }\textbf {\bibinfo {volume}
  {101}},\ \bibinfo {pages} {062308} (\bibinfo {year} {2020})}\BibitemShut
  {NoStop}%
\bibitem [{\citenamefont {Sawant}\ \emph {et~al.}(2020)\citenamefont {Sawant},
  \citenamefont {Blackmore}, \citenamefont {Gregory}, \citenamefont
  {Mur-Petit}, \citenamefont {Jaksch}, \citenamefont {Aldegunde}, \citenamefont
  {Hutson}, \citenamefont {Tarbutt},\ and\ \citenamefont
  {Cornish}}]{sawant2020ultracold}%
  \BibitemOpen
  \bibfield  {author} {\bibinfo {author} {\bibfnamefont {R.}~\bibnamefont
  {Sawant}}, \bibinfo {author} {\bibfnamefont {J.~A.}\ \bibnamefont
  {Blackmore}}, \bibinfo {author} {\bibfnamefont {P.~D.}\ \bibnamefont
  {Gregory}}, \bibinfo {author} {\bibfnamefont {J.}~\bibnamefont {Mur-Petit}},
  \bibinfo {author} {\bibfnamefont {D.}~\bibnamefont {Jaksch}}, \bibinfo
  {author} {\bibfnamefont {J.}~\bibnamefont {Aldegunde}}, \bibinfo {author}
  {\bibfnamefont {J.~M.}\ \bibnamefont {Hutson}}, \bibinfo {author}
  {\bibfnamefont {M.}~\bibnamefont {Tarbutt}}, \ and\ \bibinfo {author}
  {\bibfnamefont {S.~L.}\ \bibnamefont {Cornish}},\ }\href@noop {} {\bibfield
  {journal} {\bibinfo  {journal} {New Journal of Physics}\ }\textbf {\bibinfo
  {volume} {22}},\ \bibinfo {pages} {013027} (\bibinfo {year}
  {2020})}\BibitemShut {NoStop}%
\bibitem [{\citenamefont {Rvachov}\ \emph {et~al.}(2017)\citenamefont
  {Rvachov}, \citenamefont {Son}, \citenamefont {Sommer}, \citenamefont
  {Ebadi}, \citenamefont {Park}, \citenamefont {Zwierlein}, \citenamefont
  {Ketterle},\ and\ \citenamefont {Jamison}}]{NaLiGround}%
  \BibitemOpen
  \bibfield  {author} {\bibinfo {author} {\bibfnamefont {T.~M.}\ \bibnamefont
  {Rvachov}}, \bibinfo {author} {\bibfnamefont {H.}~\bibnamefont {Son}},
  \bibinfo {author} {\bibfnamefont {A.~T.}\ \bibnamefont {Sommer}}, \bibinfo
  {author} {\bibfnamefont {S.}~\bibnamefont {Ebadi}}, \bibinfo {author}
  {\bibfnamefont {J.~J.}\ \bibnamefont {Park}}, \bibinfo {author}
  {\bibfnamefont {M.~W.}\ \bibnamefont {Zwierlein}}, \bibinfo {author}
  {\bibfnamefont {W.}~\bibnamefont {Ketterle}}, \ and\ \bibinfo {author}
  {\bibfnamefont {A.~O.}\ \bibnamefont {Jamison}},\ }\href@noop {} {\bibfield
  {journal} {\bibinfo  {journal} {Phys. Rev. Lett.}\ }\textbf {\bibinfo
  {volume} {119}},\ \bibinfo {pages} {143001} (\bibinfo {year}
  {2017})}\BibitemShut {NoStop}%
\bibitem [{\citenamefont {Ni}\ \emph {et~al.}(2008)\citenamefont {Ni},
  \citenamefont {Ospelkaus}, \citenamefont {De~Miranda}, \citenamefont {Pe'Er},
  \citenamefont {Neyenhuis}, \citenamefont {Zirbel}, \citenamefont
  {Kotochigova}, \citenamefont {Julienne}, \citenamefont {Jin},\ and\
  \citenamefont {Ye}}]{ni2008high}%
  \BibitemOpen
  \bibfield  {author} {\bibinfo {author} {\bibfnamefont {K.-K.}\ \bibnamefont
  {Ni}}, \bibinfo {author} {\bibfnamefont {S.}~\bibnamefont {Ospelkaus}},
  \bibinfo {author} {\bibfnamefont {M.}~\bibnamefont {De~Miranda}}, \bibinfo
  {author} {\bibfnamefont {A.}~\bibnamefont {Pe'Er}}, \bibinfo {author}
  {\bibfnamefont {B.}~\bibnamefont {Neyenhuis}}, \bibinfo {author}
  {\bibfnamefont {J.}~\bibnamefont {Zirbel}}, \bibinfo {author} {\bibfnamefont
  {S.}~\bibnamefont {Kotochigova}}, \bibinfo {author} {\bibfnamefont
  {P.}~\bibnamefont {Julienne}}, \bibinfo {author} {\bibfnamefont
  {D.}~\bibnamefont {Jin}}, \ and\ \bibinfo {author} {\bibfnamefont
  {J.}~\bibnamefont {Ye}},\ }\href@noop {} {\bibfield  {journal} {\bibinfo
  {journal} {science}\ }\textbf {\bibinfo {volume} {322}},\ \bibinfo {pages}
  {231} (\bibinfo {year} {2008})}\BibitemShut {NoStop}%
\bibitem [{\citenamefont {Winkler}\ \emph {et~al.}(2007)\citenamefont
  {Winkler}, \citenamefont {Lang}, \citenamefont {Thalhammer}, \citenamefont
  {vd~Straten}, \citenamefont {Grimm},\ and\ \citenamefont
  {Denschlag}}]{winkler2007coherent}%
  \BibitemOpen
  \bibfield  {author} {\bibinfo {author} {\bibfnamefont {K.}~\bibnamefont
  {Winkler}}, \bibinfo {author} {\bibfnamefont {F.}~\bibnamefont {Lang}},
  \bibinfo {author} {\bibfnamefont {G.}~\bibnamefont {Thalhammer}}, \bibinfo
  {author} {\bibfnamefont {P.}~\bibnamefont {vd~Straten}}, \bibinfo {author}
  {\bibfnamefont {R.}~\bibnamefont {Grimm}}, \ and\ \bibinfo {author}
  {\bibfnamefont {J.~H.}\ \bibnamefont {Denschlag}},\ }\href@noop {} {\bibfield
   {journal} {\bibinfo  {journal} {Physical review letters}\ }\textbf {\bibinfo
  {volume} {98}},\ \bibinfo {pages} {043201} (\bibinfo {year}
  {2007})}\BibitemShut {NoStop}%
\bibitem [{\citenamefont {Danzl}\ \emph {et~al.}(2010)\citenamefont {Danzl},
  \citenamefont {Mark}, \citenamefont {Haller}, \citenamefont {Gustavsson},
  \citenamefont {Hart}, \citenamefont {Aldegunde}, \citenamefont {Hutson},\
  and\ \citenamefont {N{\"a}gerl}}]{danzl2010ultracold}%
  \BibitemOpen
  \bibfield  {author} {\bibinfo {author} {\bibfnamefont {J.~G.}\ \bibnamefont
  {Danzl}}, \bibinfo {author} {\bibfnamefont {M.~J.}\ \bibnamefont {Mark}},
  \bibinfo {author} {\bibfnamefont {E.}~\bibnamefont {Haller}}, \bibinfo
  {author} {\bibfnamefont {M.}~\bibnamefont {Gustavsson}}, \bibinfo {author}
  {\bibfnamefont {R.}~\bibnamefont {Hart}}, \bibinfo {author} {\bibfnamefont
  {J.}~\bibnamefont {Aldegunde}}, \bibinfo {author} {\bibfnamefont {J.~M.}\
  \bibnamefont {Hutson}}, \ and\ \bibinfo {author} {\bibfnamefont {H.-C.}\
  \bibnamefont {N{\"a}gerl}},\ }\href@noop {} {\bibfield  {journal} {\bibinfo
  {journal} {Nature Physics}\ }\textbf {\bibinfo {volume} {6}},\ \bibinfo
  {pages} {265} (\bibinfo {year} {2010})}\BibitemShut {NoStop}%
\bibitem [{\citenamefont {Park}\ \emph {et~al.}(2015)\citenamefont {Park},
  \citenamefont {Will},\ and\ \citenamefont {Zwierlein}}]{park2015ultracold}%
  \BibitemOpen
  \bibfield  {author} {\bibinfo {author} {\bibfnamefont {J.~W.}\ \bibnamefont
  {Park}}, \bibinfo {author} {\bibfnamefont {S.~A.}\ \bibnamefont {Will}}, \
  and\ \bibinfo {author} {\bibfnamefont {M.~W.}\ \bibnamefont {Zwierlein}},\
  }\href@noop {} {\bibfield  {journal} {\bibinfo  {journal} {Physical review
  letters}\ }\textbf {\bibinfo {volume} {114}},\ \bibinfo {pages} {205302}
  (\bibinfo {year} {2015})}\BibitemShut {NoStop}%
\bibitem [{\citenamefont {Danzl}\ \emph {et~al.}(2008)\citenamefont {Danzl},
  \citenamefont {Haller}, \citenamefont {Gustavsson}, \citenamefont {Mark},
  \citenamefont {Hart}, \citenamefont {Bouloufa}, \citenamefont {Dulieu},
  \citenamefont {Ritsch},\ and\ \citenamefont {N{\"a}gerl}}]{danzl2008quantum}%
  \BibitemOpen
  \bibfield  {author} {\bibinfo {author} {\bibfnamefont {J.~G.}\ \bibnamefont
  {Danzl}}, \bibinfo {author} {\bibfnamefont {E.}~\bibnamefont {Haller}},
  \bibinfo {author} {\bibfnamefont {M.}~\bibnamefont {Gustavsson}}, \bibinfo
  {author} {\bibfnamefont {M.~J.}\ \bibnamefont {Mark}}, \bibinfo {author}
  {\bibfnamefont {R.}~\bibnamefont {Hart}}, \bibinfo {author} {\bibfnamefont
  {N.}~\bibnamefont {Bouloufa}}, \bibinfo {author} {\bibfnamefont
  {O.}~\bibnamefont {Dulieu}}, \bibinfo {author} {\bibfnamefont
  {H.}~\bibnamefont {Ritsch}}, \ and\ \bibinfo {author} {\bibfnamefont {H.-C.}\
  \bibnamefont {N{\"a}gerl}},\ }\href@noop {} {\bibfield  {journal} {\bibinfo
  {journal} {Science}\ }\textbf {\bibinfo {volume} {321}},\ \bibinfo {pages}
  {1062} (\bibinfo {year} {2008})}\BibitemShut {NoStop}%
\bibitem [{\citenamefont {Krzyzewski}\ \emph {et~al.}(2015)\citenamefont
  {Krzyzewski}, \citenamefont {Akin}, \citenamefont {Dizikes}, \citenamefont
  {Morrison},\ and\ \citenamefont {Abraham}}]{PhysRevA.92.062714}%
  \BibitemOpen
  \bibfield  {author} {\bibinfo {author} {\bibfnamefont {S.~P.}\ \bibnamefont
  {Krzyzewski}}, \bibinfo {author} {\bibfnamefont {T.~G.}\ \bibnamefont
  {Akin}}, \bibinfo {author} {\bibfnamefont {J.}~\bibnamefont {Dizikes}},
  \bibinfo {author} {\bibfnamefont {M.~A.}\ \bibnamefont {Morrison}}, \ and\
  \bibinfo {author} {\bibfnamefont {E.~R.~I.}\ \bibnamefont {Abraham}},\
  }\href@noop {} {\bibfield  {journal} {\bibinfo  {journal} {Phys. Rev. A}\
  }\textbf {\bibinfo {volume} {92}},\ \bibinfo {pages} {062714} (\bibinfo
  {year} {2015})}\BibitemShut {NoStop}%
\bibitem [{\citenamefont {Shuman}\ \emph {et~al.}(2010)\citenamefont {Shuman},
  \citenamefont {Barry},\ and\ \citenamefont {DeMille}}]{shuman2010laser}%
  \BibitemOpen
  \bibfield  {author} {\bibinfo {author} {\bibfnamefont {E.~S.}\ \bibnamefont
  {Shuman}}, \bibinfo {author} {\bibfnamefont {J.~F.}\ \bibnamefont {Barry}}, \
  and\ \bibinfo {author} {\bibfnamefont {D.}~\bibnamefont {DeMille}},\
  }\href@noop {} {\bibfield  {journal} {\bibinfo  {journal} {Nature}\ }\textbf
  {\bibinfo {volume} {467}},\ \bibinfo {pages} {820} (\bibinfo {year}
  {2010})}\BibitemShut {NoStop}%
\bibitem [{\citenamefont {Anderegg}\ \emph {et~al.}(2018)\citenamefont
  {Anderegg}, \citenamefont {Augenbraun}, \citenamefont {Bao}, \citenamefont
  {Burchesky}, \citenamefont {Cheuk}, \citenamefont {Ketterle},\ and\
  \citenamefont {Doyle}}]{anderegg2018laser}%
  \BibitemOpen
  \bibfield  {author} {\bibinfo {author} {\bibfnamefont {L.}~\bibnamefont
  {Anderegg}}, \bibinfo {author} {\bibfnamefont {B.~L.}\ \bibnamefont
  {Augenbraun}}, \bibinfo {author} {\bibfnamefont {Y.}~\bibnamefont {Bao}},
  \bibinfo {author} {\bibfnamefont {S.}~\bibnamefont {Burchesky}}, \bibinfo
  {author} {\bibfnamefont {L.~W.}\ \bibnamefont {Cheuk}}, \bibinfo {author}
  {\bibfnamefont {W.}~\bibnamefont {Ketterle}}, \ and\ \bibinfo {author}
  {\bibfnamefont {J.~M.}\ \bibnamefont {Doyle}},\ }\href@noop {} {\bibfield
  {journal} {\bibinfo  {journal} {Nature Physics}\ }\textbf {\bibinfo {volume}
  {14}},\ \bibinfo {pages} {890} (\bibinfo {year} {2018})}\BibitemShut
  {NoStop}%
\bibitem [{\citenamefont {Hu}\ \emph {et~al.}(2019)\citenamefont {Hu},
  \citenamefont {Liu}, \citenamefont {Grimes}, \citenamefont {Lin},
  \citenamefont {Gheorghe}, \citenamefont {Vexiau}, \citenamefont
  {Bouloufa-Maafa}, \citenamefont {Dulieu}, \citenamefont {Rosenband},\ and\
  \citenamefont {Ni}}]{hu2019direct}%
  \BibitemOpen
  \bibfield  {author} {\bibinfo {author} {\bibfnamefont {M.-G.}\ \bibnamefont
  {Hu}}, \bibinfo {author} {\bibfnamefont {Y.}~\bibnamefont {Liu}}, \bibinfo
  {author} {\bibfnamefont {D.~D.}\ \bibnamefont {Grimes}}, \bibinfo {author}
  {\bibfnamefont {Y.-W.}\ \bibnamefont {Lin}}, \bibinfo {author} {\bibfnamefont
  {A.~H.}\ \bibnamefont {Gheorghe}}, \bibinfo {author} {\bibfnamefont
  {R.}~\bibnamefont {Vexiau}}, \bibinfo {author} {\bibfnamefont
  {N.}~\bibnamefont {Bouloufa-Maafa}}, \bibinfo {author} {\bibfnamefont
  {O.}~\bibnamefont {Dulieu}}, \bibinfo {author} {\bibfnamefont
  {T.}~\bibnamefont {Rosenband}}, \ and\ \bibinfo {author} {\bibfnamefont
  {K.-K.}\ \bibnamefont {Ni}},\ }\href@noop {} {\bibfield  {journal} {\bibinfo
  {journal} {Science}\ }\textbf {\bibinfo {volume} {366}},\ \bibinfo {pages}
  {1111} (\bibinfo {year} {2019})}\BibitemShut {NoStop}%
\bibitem [{\citenamefont {Liu}\ \emph {et~al.}(2020)\citenamefont {Liu},
  \citenamefont {Hu}, \citenamefont {Nichols}, \citenamefont {Grimes},
  \citenamefont {Karman}, \citenamefont {Guo},\ and\ \citenamefont
  {Ni}}]{liu2020photo}%
  \BibitemOpen
  \bibfield  {author} {\bibinfo {author} {\bibfnamefont {Y.}~\bibnamefont
  {Liu}}, \bibinfo {author} {\bibfnamefont {M.-G.}\ \bibnamefont {Hu}},
  \bibinfo {author} {\bibfnamefont {M.~A.}\ \bibnamefont {Nichols}}, \bibinfo
  {author} {\bibfnamefont {D.~D.}\ \bibnamefont {Grimes}}, \bibinfo {author}
  {\bibfnamefont {T.}~\bibnamefont {Karman}}, \bibinfo {author} {\bibfnamefont
  {H.}~\bibnamefont {Guo}}, \ and\ \bibinfo {author} {\bibfnamefont {K.-K.}\
  \bibnamefont {Ni}},\ }\href@noop {} {\bibfield  {journal} {\bibinfo
  {journal} {Nature Physics}\ }\textbf {\bibinfo {volume} {16}},\ \bibinfo
  {pages} {1132} (\bibinfo {year} {2020})}\BibitemShut {NoStop}%
\bibitem [{\citenamefont {Gregory}\ \emph
  {et~al.}(2021{\natexlab{a}})\citenamefont {Gregory}, \citenamefont
  {Blackmore}, \citenamefont {Bromley},\ and\ \citenamefont
  {Cornish}}]{gregory2021loss}%
  \BibitemOpen
  \bibfield  {author} {\bibinfo {author} {\bibfnamefont {P.}~\bibnamefont
  {Gregory}}, \bibinfo {author} {\bibfnamefont {J.}~\bibnamefont {Blackmore}},
  \bibinfo {author} {\bibfnamefont {S.}~\bibnamefont {Bromley}}, \ and\
  \bibinfo {author} {\bibfnamefont {S.}~\bibnamefont {Cornish}},\ }\href@noop
  {} {\bibfield  {journal} {\bibinfo  {journal} {Bulletin of the American
  Physical Society}\ }\textbf {\bibinfo {volume} {66}} (\bibinfo {year}
  {2021}{\natexlab{a}})}\BibitemShut {NoStop}%
\bibitem [{\citenamefont {Gregory}\ \emph
  {et~al.}(2021{\natexlab{b}})\citenamefont {Gregory}, \citenamefont
  {Blackmore}, \citenamefont {Fernley}, \citenamefont {Bromley}, \citenamefont
  {Hutson}, \citenamefont {Cornish} \emph {et~al.}}]{gregory2021molecule}%
  \BibitemOpen
  \bibfield  {author} {\bibinfo {author} {\bibfnamefont {P.~D.}\ \bibnamefont
  {Gregory}}, \bibinfo {author} {\bibfnamefont {J.~A.}\ \bibnamefont
  {Blackmore}}, \bibinfo {author} {\bibfnamefont {L.~M.}\ \bibnamefont
  {Fernley}}, \bibinfo {author} {\bibfnamefont {S.~L.}\ \bibnamefont
  {Bromley}}, \bibinfo {author} {\bibfnamefont {J.~M.}\ \bibnamefont {Hutson}},
  \bibinfo {author} {\bibfnamefont {S.~L.}\ \bibnamefont {Cornish}},  \emph
  {et~al.},\ }\href@noop {} {\bibfield  {journal} {\bibinfo  {journal} {New
  Journal of Physics}\ }\textbf {\bibinfo {volume} {23}},\ \bibinfo {pages}
  {125004} (\bibinfo {year} {2021}{\natexlab{b}})}\BibitemShut {NoStop}%
\bibitem [{\citenamefont {Gersema}\ \emph {et~al.}(2021)\citenamefont
  {Gersema}, \citenamefont {Voges}, \citenamefont {Meyer~zum Alten~Borgloh},
  \citenamefont {Koch}, \citenamefont {Hartmann}, \citenamefont {Zenesini},
  \citenamefont {Ospelkaus}, \citenamefont {Lin}, \citenamefont {He},\ and\
  \citenamefont {Wang}}]{ComplexesNaKNaRb}%
  \BibitemOpen
  \bibfield  {author} {\bibinfo {author} {\bibfnamefont {P.}~\bibnamefont
  {Gersema}}, \bibinfo {author} {\bibfnamefont {K.~K.}\ \bibnamefont {Voges}},
  \bibinfo {author} {\bibfnamefont {M.}~\bibnamefont {Meyer~zum
  Alten~Borgloh}}, \bibinfo {author} {\bibfnamefont {L.}~\bibnamefont {Koch}},
  \bibinfo {author} {\bibfnamefont {T.}~\bibnamefont {Hartmann}}, \bibinfo
  {author} {\bibfnamefont {A.}~\bibnamefont {Zenesini}}, \bibinfo {author}
  {\bibfnamefont {S.}~\bibnamefont {Ospelkaus}}, \bibinfo {author}
  {\bibfnamefont {J.}~\bibnamefont {Lin}}, \bibinfo {author} {\bibfnamefont
  {J.}~\bibnamefont {He}}, \ and\ \bibinfo {author} {\bibfnamefont
  {D.}~\bibnamefont {Wang}},\ }\href@noop {} {\bibfield  {journal} {\bibinfo
  {journal} {Phys. Rev. Lett.}\ }\textbf {\bibinfo {volume} {127}},\ \bibinfo
  {pages} {163401} (\bibinfo {year} {2021})}\BibitemShut {NoStop}%
\bibitem [{\citenamefont {Bause}\ \emph {et~al.}(2021)\citenamefont {Bause},
  \citenamefont {Schindewolf}, \citenamefont {Tao}, \citenamefont {Duda},
  \citenamefont {Chen}, \citenamefont {Qu{\'e}m{\'e}ner}, \citenamefont
  {Karman}, \citenamefont {Christianen}, \citenamefont {Bloch},\ and\
  \citenamefont {Luo}}]{bause2021collisions}%
  \BibitemOpen
  \bibfield  {author} {\bibinfo {author} {\bibfnamefont {R.}~\bibnamefont
  {Bause}}, \bibinfo {author} {\bibfnamefont {A.}~\bibnamefont {Schindewolf}},
  \bibinfo {author} {\bibfnamefont {R.}~\bibnamefont {Tao}}, \bibinfo {author}
  {\bibfnamefont {M.}~\bibnamefont {Duda}}, \bibinfo {author} {\bibfnamefont
  {X.-Y.}\ \bibnamefont {Chen}}, \bibinfo {author} {\bibfnamefont
  {G.}~\bibnamefont {Qu{\'e}m{\'e}ner}}, \bibinfo {author} {\bibfnamefont
  {T.}~\bibnamefont {Karman}}, \bibinfo {author} {\bibfnamefont
  {A.}~\bibnamefont {Christianen}}, \bibinfo {author} {\bibfnamefont
  {I.}~\bibnamefont {Bloch}}, \ and\ \bibinfo {author} {\bibfnamefont {X.-Y.}\
  \bibnamefont {Luo}},\ }\href@noop {} {\bibfield  {journal} {\bibinfo
  {journal} {Physical Review Research}\ }\textbf {\bibinfo {volume} {3}},\
  \bibinfo {pages} {033013} (\bibinfo {year} {2021})}\BibitemShut {NoStop}%
\bibitem [{\citenamefont {Idziaszek}\ and\ \citenamefont
  {Julienne}(2010)}]{idziaszek2010universal}%
  \BibitemOpen
  \bibfield  {author} {\bibinfo {author} {\bibfnamefont {Z.}~\bibnamefont
  {Idziaszek}}\ and\ \bibinfo {author} {\bibfnamefont {P.~S.}\ \bibnamefont
  {Julienne}},\ }\href@noop {} {\bibfield  {journal} {\bibinfo  {journal}
  {Physical review letters}\ }\textbf {\bibinfo {volume} {104}},\ \bibinfo
  {pages} {113202} (\bibinfo {year} {2010})}\BibitemShut {NoStop}%
\bibitem [{\citenamefont {Matsuda}\ \emph {et~al.}(2020)\citenamefont
  {Matsuda}, \citenamefont {De~Marco}, \citenamefont {Li}, \citenamefont
  {Tobias}, \citenamefont {Valtolina}, \citenamefont {Qu{\'e}m{\'e}ner},\ and\
  \citenamefont {Ye}}]{matsuda2020resonant}%
  \BibitemOpen
  \bibfield  {author} {\bibinfo {author} {\bibfnamefont {K.}~\bibnamefont
  {Matsuda}}, \bibinfo {author} {\bibfnamefont {L.}~\bibnamefont {De~Marco}},
  \bibinfo {author} {\bibfnamefont {J.-R.}\ \bibnamefont {Li}}, \bibinfo
  {author} {\bibfnamefont {W.~G.}\ \bibnamefont {Tobias}}, \bibinfo {author}
  {\bibfnamefont {G.}~\bibnamefont {Valtolina}}, \bibinfo {author}
  {\bibfnamefont {G.}~\bibnamefont {Qu{\'e}m{\'e}ner}}, \ and\ \bibinfo
  {author} {\bibfnamefont {J.}~\bibnamefont {Ye}},\ }\href@noop {} {\bibfield
  {journal} {\bibinfo  {journal} {Science}\ }\textbf {\bibinfo {volume}
  {370}},\ \bibinfo {pages} {1324} (\bibinfo {year} {2020})}\BibitemShut
  {NoStop}%
\bibitem [{\citenamefont {Schindewolf}\ \emph {et~al.}(2022)\citenamefont
  {Schindewolf}, \citenamefont {Bause}, \citenamefont {Chen}, \citenamefont
  {Duda}, \citenamefont {Karman}, \citenamefont {Bloch},\ and\ \citenamefont
  {Luo}}]{schindewolf2022evaporation}%
  \BibitemOpen
  \bibfield  {author} {\bibinfo {author} {\bibfnamefont {A.}~\bibnamefont
  {Schindewolf}}, \bibinfo {author} {\bibfnamefont {R.}~\bibnamefont {Bause}},
  \bibinfo {author} {\bibfnamefont {X.-Y.}\ \bibnamefont {Chen}}, \bibinfo
  {author} {\bibfnamefont {M.}~\bibnamefont {Duda}}, \bibinfo {author}
  {\bibfnamefont {T.}~\bibnamefont {Karman}}, \bibinfo {author} {\bibfnamefont
  {I.}~\bibnamefont {Bloch}}, \ and\ \bibinfo {author} {\bibfnamefont {X.-Y.}\
  \bibnamefont {Luo}},\ }\href@noop {} {\bibfield  {journal} {\bibinfo
  {journal} {Nature}\ }\textbf {\bibinfo {volume} {607}},\ \bibinfo {pages}
  {677} (\bibinfo {year} {2022})}\BibitemShut {NoStop}%
\bibitem [{\citenamefont {Anderegg}\ \emph {et~al.}(2021)\citenamefont
  {Anderegg}, \citenamefont {Burchesky}, \citenamefont {Bao}, \citenamefont
  {Yu}, \citenamefont {Karman}, \citenamefont {Chae}, \citenamefont {Ni},
  \citenamefont {Ketterle},\ and\ \citenamefont
  {Doyle}}]{anderegg2021observation}%
  \BibitemOpen
  \bibfield  {author} {\bibinfo {author} {\bibfnamefont {L.}~\bibnamefont
  {Anderegg}}, \bibinfo {author} {\bibfnamefont {S.}~\bibnamefont {Burchesky}},
  \bibinfo {author} {\bibfnamefont {Y.}~\bibnamefont {Bao}}, \bibinfo {author}
  {\bibfnamefont {S.~S.}\ \bibnamefont {Yu}}, \bibinfo {author} {\bibfnamefont
  {T.}~\bibnamefont {Karman}}, \bibinfo {author} {\bibfnamefont
  {E.}~\bibnamefont {Chae}}, \bibinfo {author} {\bibfnamefont {K.-K.}\
  \bibnamefont {Ni}}, \bibinfo {author} {\bibfnamefont {W.}~\bibnamefont
  {Ketterle}}, \ and\ \bibinfo {author} {\bibfnamefont {J.~M.}\ \bibnamefont
  {Doyle}},\ }\href@noop {} {\bibfield  {journal} {\bibinfo  {journal}
  {Science}\ }\textbf {\bibinfo {volume} {373}},\ \bibinfo {pages} {779}
  (\bibinfo {year} {2021})}\BibitemShut {NoStop}%
\bibitem [{\citenamefont {Wigner}(1948)}]{wigner1948behavior}%
  \BibitemOpen
  \bibfield  {author} {\bibinfo {author} {\bibfnamefont {E.~P.}\ \bibnamefont
  {Wigner}},\ }\href@noop {} {\bibfield  {journal} {\bibinfo  {journal}
  {Physical Review}\ }\textbf {\bibinfo {volume} {73}},\ \bibinfo {pages}
  {1002} (\bibinfo {year} {1948})}\BibitemShut {NoStop}%
\bibitem [{\citenamefont {Idziaszek}\ \emph {et~al.}(2015)\citenamefont
  {Idziaszek}, \citenamefont {Jachymski},\ and\ \citenamefont
  {Julienne}}]{idziaszek2015reactive}%
  \BibitemOpen
  \bibfield  {author} {\bibinfo {author} {\bibfnamefont {Z.}~\bibnamefont
  {Idziaszek}}, \bibinfo {author} {\bibfnamefont {K.}~\bibnamefont
  {Jachymski}}, \ and\ \bibinfo {author} {\bibfnamefont {P.~S.}\ \bibnamefont
  {Julienne}},\ }\href@noop {} {\bibfield  {journal} {\bibinfo  {journal} {New
  Journal of Physics}\ }\textbf {\bibinfo {volume} {17}},\ \bibinfo {pages}
  {035007} (\bibinfo {year} {2015})}\BibitemShut {NoStop}%
\bibitem [{\citenamefont {Derevianko}\ \emph {et~al.}(2001)\citenamefont
  {Derevianko}, \citenamefont {Babb},\ and\ \citenamefont
  {Dalgarno}}]{derevianko2001high}%
  \BibitemOpen
  \bibfield  {author} {\bibinfo {author} {\bibfnamefont {A.}~\bibnamefont
  {Derevianko}}, \bibinfo {author} {\bibfnamefont {J.}~\bibnamefont {Babb}}, \
  and\ \bibinfo {author} {\bibfnamefont {A.}~\bibnamefont {Dalgarno}},\
  }\href@noop {} {\bibfield  {journal} {\bibinfo  {journal} {Physical Review
  A}\ }\textbf {\bibinfo {volume} {63}},\ \bibinfo {pages} {052704} (\bibinfo
  {year} {2001})}\BibitemShut {NoStop}%
\bibitem [{\citenamefont {Chevy}\ \emph {et~al.}(2005)\citenamefont {Chevy},
  \citenamefont {Van~Kempen}, \citenamefont {Bourdel}, \citenamefont {Zhang},
  \citenamefont {Khaykovich}, \citenamefont {Teichmann}, \citenamefont
  {Tarruell}, \citenamefont {Kokkelmans},\ and\ \citenamefont
  {Salomon}}]{chevy2005resonant}%
  \BibitemOpen
  \bibfield  {author} {\bibinfo {author} {\bibfnamefont {F.}~\bibnamefont
  {Chevy}}, \bibinfo {author} {\bibfnamefont {E.}~\bibnamefont {Van~Kempen}},
  \bibinfo {author} {\bibfnamefont {T.}~\bibnamefont {Bourdel}}, \bibinfo
  {author} {\bibfnamefont {J.}~\bibnamefont {Zhang}}, \bibinfo {author}
  {\bibfnamefont {L.}~\bibnamefont {Khaykovich}}, \bibinfo {author}
  {\bibfnamefont {M.}~\bibnamefont {Teichmann}}, \bibinfo {author}
  {\bibfnamefont {L.}~\bibnamefont {Tarruell}}, \bibinfo {author}
  {\bibfnamefont {S.}~\bibnamefont {Kokkelmans}}, \ and\ \bibinfo {author}
  {\bibfnamefont {C.}~\bibnamefont {Salomon}},\ }\href@noop {} {\bibfield
  {journal} {\bibinfo  {journal} {Physical Review A}\ }\textbf {\bibinfo
  {volume} {71}},\ \bibinfo {pages} {062710} (\bibinfo {year}
  {2005})}\BibitemShut {NoStop}%
\bibitem [{\citenamefont {Bai}\ \emph {et~al.}(2019)\citenamefont {Bai},
  \citenamefont {Li}, \citenamefont {Wang},\ and\ \citenamefont
  {Cong}}]{bai2019model}%
  \BibitemOpen
  \bibfield  {author} {\bibinfo {author} {\bibfnamefont {Y.-P.}\ \bibnamefont
  {Bai}}, \bibinfo {author} {\bibfnamefont {J.-L.}\ \bibnamefont {Li}},
  \bibinfo {author} {\bibfnamefont {G.-R.}\ \bibnamefont {Wang}}, \ and\
  \bibinfo {author} {\bibfnamefont {S.-L.}\ \bibnamefont {Cong}},\ }\href@noop
  {} {\bibfield  {journal} {\bibinfo  {journal} {Physical Review A}\ }\textbf
  {\bibinfo {volume} {100}},\ \bibinfo {pages} {012705} (\bibinfo {year}
  {2019})}\BibitemShut {NoStop}%
\bibitem [{\citenamefont {Tomza}\ \emph {et~al.}(2013)\citenamefont {Tomza},
  \citenamefont {Madison}, \citenamefont {Moszynski},\ and\ \citenamefont
  {Krems}}]{TripletReactions}%
  \BibitemOpen
  \bibfield  {author} {\bibinfo {author} {\bibfnamefont {M.}~\bibnamefont
  {Tomza}}, \bibinfo {author} {\bibfnamefont {K.~W.}\ \bibnamefont {Madison}},
  \bibinfo {author} {\bibfnamefont {R.}~\bibnamefont {Moszynski}}, \ and\
  \bibinfo {author} {\bibfnamefont {R.~V.}\ \bibnamefont {Krems}},\ }\href@noop
  {} {\bibfield  {journal} {\bibinfo  {journal} {Phys. Rev. A}\ }\textbf
  {\bibinfo {volume} {88}},\ \bibinfo {pages} {050701} (\bibinfo {year}
  {2013})}\BibitemShut {NoStop}%
\bibitem [{\citenamefont {Son}\ \emph {et~al.}(2020)\citenamefont {Son},
  \citenamefont {Park}, \citenamefont {Ketterle},\ and\ \citenamefont
  {Jamison}}]{NaLiSympCool}%
  \BibitemOpen
  \bibfield  {author} {\bibinfo {author} {\bibfnamefont {H.}~\bibnamefont
  {Son}}, \bibinfo {author} {\bibfnamefont {J.~J.}\ \bibnamefont {Park}},
  \bibinfo {author} {\bibfnamefont {W.}~\bibnamefont {Ketterle}}, \ and\
  \bibinfo {author} {\bibfnamefont {A.~O.}\ \bibnamefont {Jamison}},\
  }\href@noop {} {\bibfield  {journal} {\bibinfo  {journal} {Nature}\ }\textbf
  {\bibinfo {volume} {580}},\ \bibinfo {pages} {197} (\bibinfo {year}
  {2020})}\BibitemShut {NoStop}%
\bibitem [{\citenamefont {De~Marco}\ \emph {et~al.}(2019)\citenamefont
  {De~Marco}, \citenamefont {Valtolina}, \citenamefont {Matsuda}, \citenamefont
  {Tobias}, \citenamefont {Covey},\ and\ \citenamefont
  {Ye}}]{de2019degenerate}%
  \BibitemOpen
  \bibfield  {author} {\bibinfo {author} {\bibfnamefont {L.}~\bibnamefont
  {De~Marco}}, \bibinfo {author} {\bibfnamefont {G.}~\bibnamefont {Valtolina}},
  \bibinfo {author} {\bibfnamefont {K.}~\bibnamefont {Matsuda}}, \bibinfo
  {author} {\bibfnamefont {W.~G.}\ \bibnamefont {Tobias}}, \bibinfo {author}
  {\bibfnamefont {J.~P.}\ \bibnamefont {Covey}}, \ and\ \bibinfo {author}
  {\bibfnamefont {J.}~\bibnamefont {Ye}},\ }\href@noop {} {\bibfield  {journal}
  {\bibinfo  {journal} {Science}\ }\textbf {\bibinfo {volume} {363}},\ \bibinfo
  {pages} {853} (\bibinfo {year} {2019})}\BibitemShut {NoStop}%
\bibitem [{\citenamefont {Mies}\ \emph {et~al.}(1996)\citenamefont {Mies},
  \citenamefont {Williams}, \citenamefont {Julienne},\ and\ \citenamefont
  {Krauss}}]{Julienne:96}%
  \BibitemOpen
  \bibfield  {author} {\bibinfo {author} {\bibfnamefont {F.~H.}\ \bibnamefont
  {Mies}}, \bibinfo {author} {\bibfnamefont {C.~J.}\ \bibnamefont {Williams}},
  \bibinfo {author} {\bibfnamefont {P.~S.}\ \bibnamefont {Julienne}}, \ and\
  \bibinfo {author} {\bibfnamefont {M.}~\bibnamefont {Krauss}},\ }\href@noop {}
  {\bibfield  {journal} {\bibinfo  {journal} {Journal of research of the
  National Institute of Standards and Technology}\ }\textbf {\bibinfo {volume}
  {101}},\ \bibinfo {pages} {521} (\bibinfo {year} {1996})}\BibitemShut
  {NoStop}%
\bibitem [{\citenamefont {Krems}\ and\ \citenamefont
  {Dalgarno}(2004)}]{Krems:04}%
  \BibitemOpen
  \bibfield  {author} {\bibinfo {author} {\bibfnamefont {R.}~\bibnamefont
  {Krems}}\ and\ \bibinfo {author} {\bibfnamefont {A.}~\bibnamefont
  {Dalgarno}},\ }\href@noop {} {\bibfield  {journal} {\bibinfo  {journal} {The
  Journal of chemical physics}\ }\textbf {\bibinfo {volume} {120}},\ \bibinfo
  {pages} {2296} (\bibinfo {year} {2004})}\BibitemShut {NoStop}%
\bibitem [{\citenamefont {Tscherbul}\ \emph {et~al.}(2009)\citenamefont
  {Tscherbul}, \citenamefont {Suleimanov}, \citenamefont {Aquilanti},\ and\
  \citenamefont {Krems}}]{Tscherbul:09}%
  \BibitemOpen
  \bibfield  {author} {\bibinfo {author} {\bibfnamefont {T.}~\bibnamefont
  {Tscherbul}}, \bibinfo {author} {\bibfnamefont {Y.~V.}\ \bibnamefont
  {Suleimanov}}, \bibinfo {author} {\bibfnamefont {V.}~\bibnamefont
  {Aquilanti}}, \ and\ \bibinfo {author} {\bibfnamefont {R.}~\bibnamefont
  {Krems}},\ }\href@noop {} {\bibfield  {journal} {\bibinfo  {journal} {New
  Journal of Physics}\ }\textbf {\bibinfo {volume} {11}},\ \bibinfo {pages}
  {055021} (\bibinfo {year} {2009})}\BibitemShut {NoStop}%
\bibitem [{\citenamefont {Stone}(2013)}]{StoneBook}%
  \BibitemOpen
  \bibfield  {author} {\bibinfo {author} {\bibfnamefont {A.}~\bibnamefont
  {Stone}},\ }\href@noop {} {\emph {\bibinfo {title} {The Theory of
  Intermolecular Forces}}}\ (\bibinfo  {publisher} {Oxford University Press},\
  \bibinfo {address} {Oxford},\ \bibinfo {year} {2013})\ Chap.\ \bibinfo
  {chapter} {3.3}\BibitemShut {NoStop}%
\bibitem [{\citenamefont {Gronowski}\ \emph {et~al.}(2020)\citenamefont
  {Gronowski}, \citenamefont {Koza},\ and\ \citenamefont
  {Tomza}}]{Gronowski:20}%
  \BibitemOpen
  \bibfield  {author} {\bibinfo {author} {\bibfnamefont {M.}~\bibnamefont
  {Gronowski}}, \bibinfo {author} {\bibfnamefont {A.~M.}\ \bibnamefont {Koza}},
  \ and\ \bibinfo {author} {\bibfnamefont {M.}~\bibnamefont {Tomza}},\
  }\href@noop {} {\bibfield  {journal} {\bibinfo  {journal} {Physical Review
  A}\ }\textbf {\bibinfo {volume} {102}},\ \bibinfo {pages} {020801} (\bibinfo
  {year} {2020})}\BibitemShut {NoStop}%
\bibitem [{\citenamefont {Harrison}\ and\ \citenamefont
  {Lawson}(2005)}]{Harrison:05}%
  \BibitemOpen
  \bibfield  {author} {\bibinfo {author} {\bibfnamefont {J.~F.}\ \bibnamefont
  {Harrison}}\ and\ \bibinfo {author} {\bibfnamefont {D.~B.}\ \bibnamefont
  {Lawson}},\ }\href@noop {} {\bibfield  {journal} {\bibinfo  {journal}
  {International journal of quantum chemistry}\ }\textbf {\bibinfo {volume}
  {102}},\ \bibinfo {pages} {1087} (\bibinfo {year} {2005})}\BibitemShut
  {NoStop}%
\bibitem [{\citenamefont {Hermsmeier}\ \emph {et~al.}(2021)\citenamefont
  {Hermsmeier}, \citenamefont {K{\l}os}, \citenamefont {Kotochigova},\ and\
  \citenamefont {Tscherbul}}]{Hermsmeier:21}%
  \BibitemOpen
  \bibfield  {author} {\bibinfo {author} {\bibfnamefont {R.}~\bibnamefont
  {Hermsmeier}}, \bibinfo {author} {\bibfnamefont {J.}~\bibnamefont {K{\l}os}},
  \bibinfo {author} {\bibfnamefont {S.}~\bibnamefont {Kotochigova}}, \ and\
  \bibinfo {author} {\bibfnamefont {T.~V.}\ \bibnamefont {Tscherbul}},\
  }\href@noop {} {\bibfield  {journal} {\bibinfo  {journal} {Physical review
  letters}\ }\textbf {\bibinfo {volume} {127}},\ \bibinfo {pages} {103402}
  (\bibinfo {year} {2021})}\BibitemShut {NoStop}%
\bibitem [{\citenamefont {Ismail}\ \emph {et~al.}(2016)\citenamefont {Ismail},
  \citenamefont {Kores}, \citenamefont {Geskus},\ and\ \citenamefont
  {Pollnau}}]{ismail2016fabry}%
  \BibitemOpen
  \bibfield  {author} {\bibinfo {author} {\bibfnamefont {N.}~\bibnamefont
  {Ismail}}, \bibinfo {author} {\bibfnamefont {C.~C.}\ \bibnamefont {Kores}},
  \bibinfo {author} {\bibfnamefont {D.}~\bibnamefont {Geskus}}, \ and\ \bibinfo
  {author} {\bibfnamefont {M.}~\bibnamefont {Pollnau}},\ }\href@noop {}
  {\bibfield  {journal} {\bibinfo  {journal} {Optics express}\ }\textbf
  {\bibinfo {volume} {24}},\ \bibinfo {pages} {16366} (\bibinfo {year}
  {2016})}\BibitemShut {NoStop}%
\bibitem [{\citenamefont {Friedrich}(2017)}]{FriedrichBook}%
  \BibitemOpen
  \bibfield  {author} {\bibinfo {author} {\bibfnamefont {H.}~\bibnamefont
  {Friedrich}},\ }\href@noop {} {\emph {\bibinfo {title} {Theoretical Atomic
  Physics}}},\ \bibinfo {edition} {4th}\ ed.\ (\bibinfo  {publisher} {Springer
  International},\ \bibinfo {address} {New York},\ \bibinfo {year}
  {2017})\BibitemShut {NoStop}%
\end{thebibliography}%

\clearpage

\section{Methods}

\subsection{Experimental sequence}

Our experiments use ground state $^{23}\text{Na}^{6}\text{Li}$ $ (a^{3}\Sigma ^{+})$ molecules in their lower stretched hyperfine state  ($\ket{F, M_F} = \ket{7/2, -7/2}$), where all nuclear and electron spins are anti-aligned to the bias magnetic field direction, trapped in a 1-dimensional (1D) optical lattice made with 1596 nm light. As described previously \cite{NaLiSympCool, NaLiGround, son2022control}, we first produce loosely bound molecules via magnetic association at a Feshbach resonance near $745$ G followed by stimulated Raman adiabatic passage (STIRAP) to the rovibrational ground state. These triplet ground state molecules are in the upper stretched hyperfine ($\ket{F, M_F} = \ket{7/2, 7/2}$). The bias field is dropped from $745 $ G to a low field near $8 $G in $15 $ms where a magnetic field sweep in the presence of radiofrequency waves coherently transfers them from the upper stretched state to the lower stretched state. 

After state preparation, the magnetic field is ramped to a target value in $15$ ms. A search for scattering resonances is done for the bias field range of $40.5$ G $<B<$ $1401.6$ G. After waiting a certain time for the molecules to collide with one another at the target field, we drop the field back to 8 G for reverse state transfer ($\ket{7/2, -7/2} \rightarrow  \ket{7/2, 7/2}$). The field is raised back to 745 G where the molecules are dissociated. We use absorption imaging of the resulting lithium atoms to measure molecule number and temperature. A hold time of $15$ ms $<$ t $<$ $30$ ms after each magnetic field ramp was sufficient for the bias field to settle within the range of the magnetic field inhomogeneity across the molecular sample. 


\subsection{Model for molecular decay \& density calibration}

\indent We model two-body loss with a differential equation that takes the time dependence of temperature into account: \cite{de2019degenerate}

\begin{equation}    
    \begin{aligned}
        & \dot{n}(t) = -\beta(T(t)) n^{2}(t) - \frac{3}{2}n(t)\frac{\dot{T}(t)}{T(t)} , \\
	\end{aligned}
	 \label{eq:diffeqn}
\end{equation}   
where $\beta$ is the two-body loss rate constant, $n$ is the mean density, and $T$ is the temperature of the molecules. 
Molecules are lost preferentially from the highest density region.  This ``anti-evaporation'' causes temperature increases of up to 50\% within one molecular decay time near $334.9$ G. We fit measured temperatures to a linear function of time, $T(t)=Ht+T_{o}$, where $H$ is the heating rate and $T_{o}$ is the initial temperature. Both away from and near to the resonance, the loss rate coefficient has a temperature dependence that can be expressed as $\beta=\beta_{0}(T(t)/T_{0})$, where $\beta_{0}$ is the initial loss rate coefficient when the temperature is $T_{0}$. To determine the rate coefficient from Eq.~(\ref{eq:diffeqn}) requires accurate knowledge of the molecular density. The mean molecular density can be expressed with the effective number of particles, $N^{\text{eff}}$, and the mean volume, $V_{\text{eff}}$, of molecules for a single pancake as $n=N^{\text{eff}}/V_{\text{eff}}$.

\indent We obtained the effective particle numbers for a single pancake from the measured number of molecules, $N^{\text{tot}}$, and the number distribution over pancakes. The observed axial profile of NaLi follows a Gaussian form with width $\sigma{=}450(60) \mu\text{m}$, so we assume a Gaussian distribution of the particle number per pancake. As the average weighted over a Gaussian, the effective particle number per pancake is $N^{\text{eff}} {=} N^{\text{tot}}\cdot a/(2\sqrt{\pi}\cdot \sigma )$, where the lattice constant, $a{=}\lambda/2$ and $\lambda {=} 1596$ nm. 

\indent The trap volume of each pancake, $V_{\text{eff}}$, is determined from the measured molecule temperature and trap frequencies. For a purely harmonic trap one obtains $V_{\text{eff}}^{(0)} = \bar{\omega}^{-3}(4\pi k_B T/m)^{3/2}$ where the geometric mean of the NaLi trap frequencies, $\bar{\omega} {=} (\omega_x\omega_y\omega_z)^{1/3}$. 
However, there are two corrections which we determine separately:  (1) the confinement in each pancake is moderately anharmonic, and (2) the system is in the cross-over regime between quasi-2D and 3D, $k_B T \; \sim \; \hbar \omega_z $.

First, the anharmonicity of the trapping potential leads to a modified mean volume $V_{\text{eff}}^{(1)}$ 

\begin{equation}
   1/V_{\text{eff}}^{(1)} = \cfrac{\int dV \text{e}^{-2\beta U(r, z)}}{[\int dV \text{e}^{-\beta U(r, z)}]^2}
    \label{eq:volume 2}
\end{equation}
where $U(r,z)$ is the potential of a single lattice site, $r$ is the radial coordinate, $z$ is the axial coordinate along the beam direction, and $\beta = (k_B T)^{-1}$. We use the same trap model validated in ref. \cite{son2022control} to determine $U(r,z)$ and the same integration limits. With typical conditions for molecular loss measurements, the mean volume $V_{\text{eff}}^{(1)}$ is larger than $V_{\text{eff}}^{(0)}$ by less than $25\%$. However, some  measurements at lower density  required  weaker optical traps for which the anharmonicity correction is larger and had to be taken into account for proper density calibration.

Second, the tight confinement in the lattice direction makes the classical thermal distribution for harmonic trapping in the pancakes invalid for low molecular temperatures. We estimate the corrected volume as

\begin{equation}
   1/V_{\text{eff}}^{(2)} =\cfrac{\int dV ( \sum_{i=0}^{\infty} \rho(r)\lvert \phi_i(z) \rvert^2 \text{e}^{\beta(-i-\frac{1}{2})\hbar \omega_z})^2}{[\int dV \sum_{i=0}^{\infty} \rho(r)\lvert \phi_i(z) \rvert^2 \text{e}^{\beta(-i-\frac{1}{2})\hbar \omega_z}]^2} 
    \label{eq:2}
\end{equation}
where $\phi_i(z)$ is the $i^{\rm th}$ eigenstate of the axial harmonic oscillator, and $\rho(r)$ is the classical thermal distribution in the radial direction, which is a Gaussian function. We confirmed that for all the experimental conditions $V_{\text{eff}}^{(0)}$ differs from the more accurate $V_{\text{eff}}^{(2)}$ by less than $20\%$. Since it is unclear how this correction might interact with the larger correction from $V_{\text{eff}}^{(1)}$, we include $V_{\text{eff}}^{(2)}$ only as an enlargement in the uncertainty.

\subsection{Hyperfine structure of NaLi}

\begin{figure}
    \centering
	\includegraphics[width = 83mm, keepaspectratio]{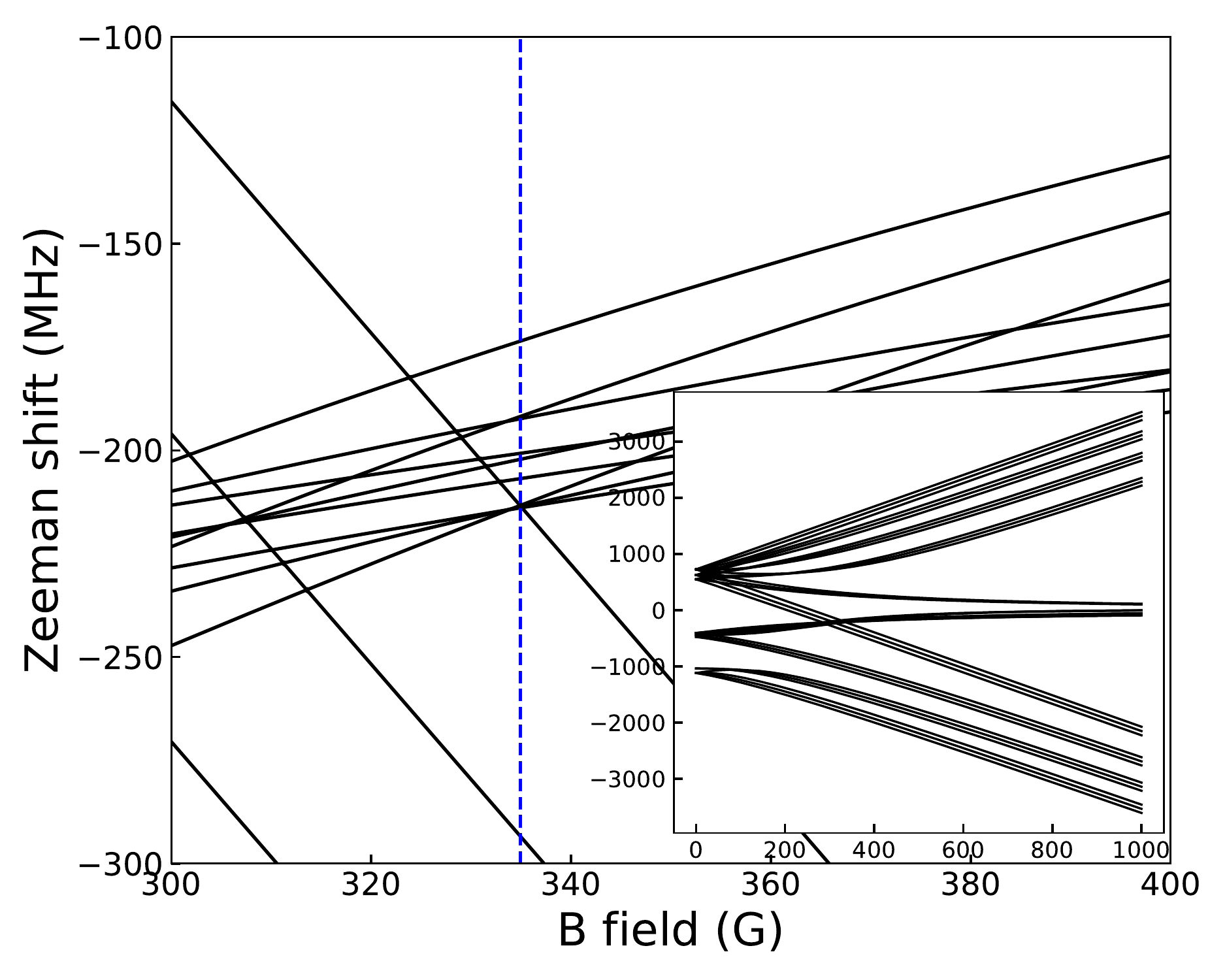}
	\caption{\textbf{Hyperfine structure of NaLi(ground) in an external magnetic field.} The blue dashed vertical line indicates the position of the Feshbach resonance ($\sim 334.92 \text{ G}$). The subplot shows the Zeeman energies of NaLi hyperfine states from 0 to 1000 G whereas the main plot is zoomed into where there are 9 near degenerate hyperfine states (between 300 to 400 G). }
	\label{fig:zeeman shift}
\end{figure}

\begin{figure}
    \centering
	\includegraphics[width = 83mm, keepaspectratio]{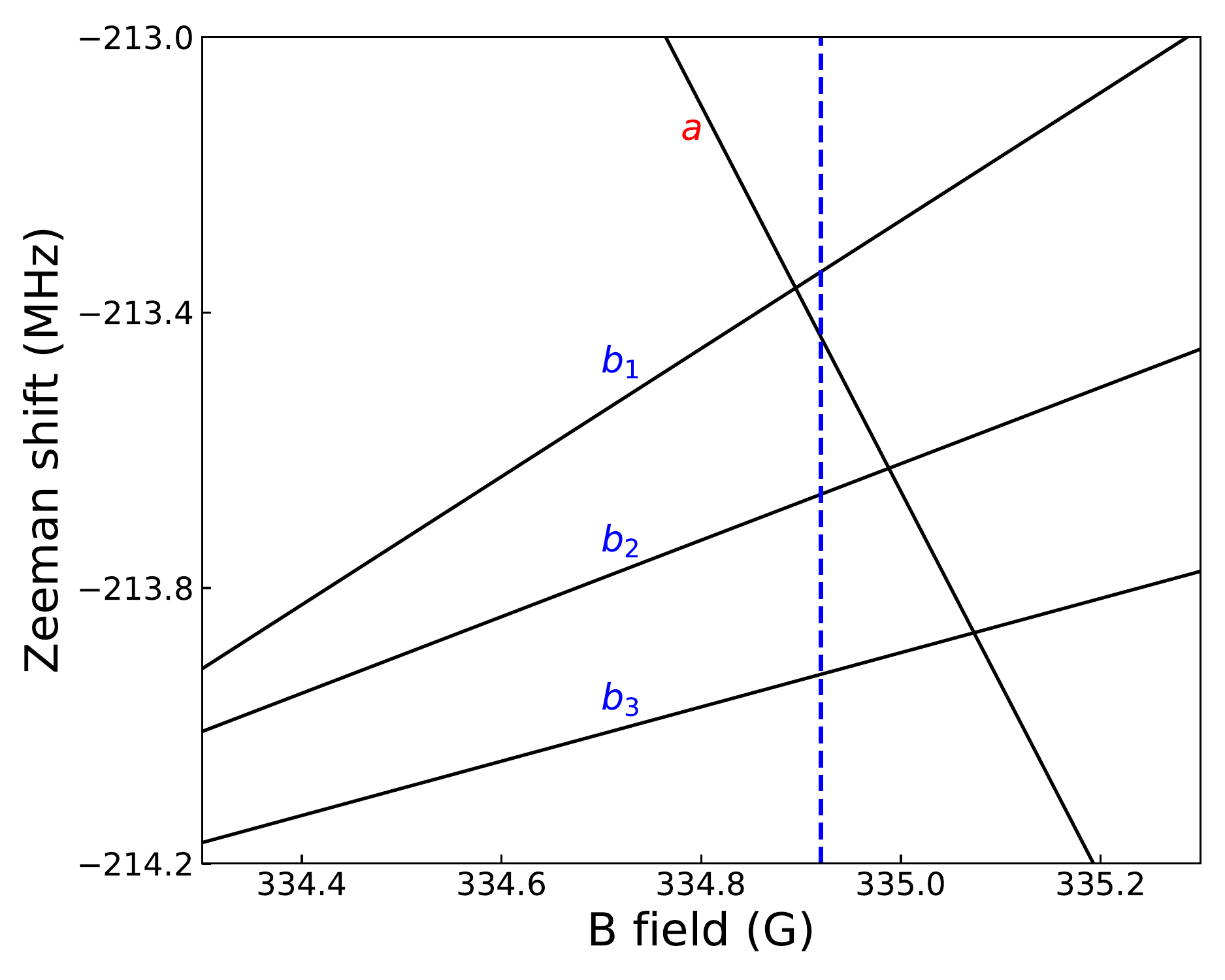}
	\caption{\textbf{Hyperfine structure of NaLi(ground) in an external magnetic field.}  State $a$ in red is the lower stretched hyperfine state of NaLi molecules. States $b_1$, $b_2$, and $b_3$ in blue are other hyperfine states that are energetically close to state $a$ near $334.92$ G.  }
	\label{fig:zeeman shift 2}
\end{figure}

The Hamiltonian that includes hyperfine coupling and Zeeman energy is given as

\begin{equation}
\begin{split}
 H & = H_{\rm HF} + H_{\rm Zeeman}  \\
    & = a_{1} \vec{S}\cdot \vec{I}_{\rm Na} +a_{2} \vec{S}\cdot \vec{I}_{\rm Li} \\ 
    & + \frac{\mu_{B}}{\hbar}(g_s \vec{S} +g_{I_{\rm Na}} \vec{I}_{\rm Na} + g_{I_{\rm Li}} \vec{I}_{\rm Li} )\cdot \vec{B} \\
\end{split}
\label{eq:Hamiltonian}
\end{equation}
where $a_1 = 433.2(1)$ MHz and $a_2 = 74.6(1)$ MHz \cite{NaLiGround}. There are 36 states in the ground rotational state manifold ($N = 0$). The states
converge to 8 hyperfine thresholds in the zero-field limit due to the conservation of the total angular momentum $ F = \lvert \vec{F} \rvert = \lvert \vec{S} + \vec{I}_{\rm{Na}} + \vec{I}_{\rm{Li}}\rvert$, where $S = \lvert \vec{S} \rvert = 1$ is the total electron spin of NaLi($a^3\Sigma^+$) and $I_{\text{Na}} = 3/2$ and $I_{\rm{Li}} = 1$ are the nuclear spins of Na and Li. The hyperfine splitting due to the Na nucleus is significantly larger than that due to the Li nucleus, so $F_1 = \lvert \vec{S} + \vec{I}_{\rm{Na}} \rvert = 1/2, 3/2$, and $5/2$ is an approximately good quantum number that characterizes the largest-scale hyperfine splittings in the zero-field limit in Fig. \ref{fig:zeeman shift}. 
The hyperfine structure of NaLi in an external magnetic field is obtained by diagonalizing the Hamiltonian given in Eq. (\ref{eq:Hamiltonian}). A total of 9 hyperfine states cross the lower stretched state at bias fields between $320$ G and $340$ G. There are three states that cross at fields within a few hundred mG of the resonance. Due to the uncertainty of the hyperfine constants, the uncertainty of their energies relative to the stretched state is $ \pm 400$ kHz, and therefore each of them is a possible candidate to cross the stretched state at or near the observed Feshbach resonance at $334.92$ G.  

The state indicated with an ``a" in Fig. \ref{fig:zeeman shift 2}  is the lower stretched state ($\ket{F=7/2,M_F=-7/2}$) which has the spin character $\ket{M_s,M_{I_{\rm{Na}}},M_{I_{\rm{Li}}}}=\ket{-1, -3/2, -1}$. The three states close in energy near the resonance are indicated as $b_1,b_2$ and $b_3$ in Fig. \ref{fig:zeeman shift 2}. The most likely candidate to couple to the stretched state is the state $b_1 {=} \ket{F=5/2,M_F=-3/2}$ since its nuclear spin character overlaps with the stretched state by 50\%, whereas for the other two states, in leading order, the spin overlap is zero. Non-zero nuclear spin overlap less than 0.1\% may arise from small intra-molecular spin-spin 
and spin-rotation 
couplings. More explicitly, the state $b_1$ has spin contributions of $0.503$ of $\ket{1, -3/2, -1}$, $0.324$ of $\ket{0, -1/2, -1}$, $0.163$ of $\ket{-1, 1/2, -1}$, etc. State $b_2 {=} \ket{F=5/2,M_F=-1/2}$ has $0.583$ of $\ket{0, 1/2, -1}$, $0.301$ of $\ket{1, -1/2, -1}$, $0.097$ of $\ket{-1, 3/2, -1}$, and some other minor contributions. State $b_3 {=} \ket{F=5/2,M_F=1/2}$ has $0.837$ of $\ket{0, 3/2, -1}$, $0.145$ of $\ket{1, 1/2, 0}$, and some other minor contributions. The difference in $M_F$ to the lower stretched state is the smallest for state $b_1$ which is 2.

\subsection{Long-range interactions of NaLi(a$^3\Sigma^+$) molecules}

Here we show that the \emph{long-range} coupling between the two hyperfine states $\ket{a}$ and $\ket{b_1}$ involved in the crossing shown in Fig.~\ref{fig:zeeman shift 2} is too weak to explain the observed loss rates. To explain the observed Feshbach resonance, it is therefore necessary to assume the presence of a \emph{short-range} loss mechanism. A minimal model for such a mechanism involves a bound state (channel $\ket{3}$) coupled to the open channels ($\ket{aa}$ and $\ket{ab_1}$).

Long-range coupling would occur outside the centrifugal $p$-wave barrier of the input channel. The peak of the barrier is at $24^{1/4} \cdot R_\text{vdw} \approx 7.8\;{\rm nm}$ where the van der Waals length $R_{\text{vdw}}=\frac{1}{2}\left(\frac{2\mu C_{6}}{\hbar^2} \right)^{1/4}$, and the inner turning point is at 100 nm at $3.35\;{\rm \mu K}$ temperature.
In the Born approximation, for a potential of average strength $V_0$ in an effective volume $4 \pi R_0^3/3$, the total low energy elastic scattering cross section 
 $\sigma = 4 \pi ( \frac{2 \mu V_0 R_0^3 }{3 \hbar^2})^2$ (identical to the solution for a spherical square well potential with radius $R_0$). 
Applying this relation to the observed nearly unitarity limited cross section of $2.95\times 10^{-11}$ cm$^2$ corresponding to the loss rate constant of $10^{-10}\;{\rm cm^3 s^{-1}}$ at 1~$\mu$K and using the position of the inner turning point to estimate $R_0=100$~nm, provides a coupling matrix element $V_0$ of 16~kHz. This is the required value for coupling outside the $p$-wave barrier to be compatible with the observed loss rates. For inelastic collisions with final wavevector $k'$, the rate has an additional factor $k'/k$ due to the density of states, but for large $k'$, the matrix element will decrease  with $k'$, so our rough estimate for the required spatial coupling matrix element should also apply to inelastic collisions.

First, we show that magnetic dipolar interactions \cite{Julienne:96} which lead to spin exchange and dipolar relaxation and often limit the lifetime of magnetically trapped atoms, are very weak outside the barrier. At the position of the $p$-wave barrier ($R_b=100$~nm), the interaction between two spins with magnetic moments $2\mu_0$, where $\mu_0$ is the Bohr magneton, is $V^{mDD}=0.052$~kHz, which is already small.  However, due to the selection rules of the magnetic dipolar interaction ($\lvert \Delta M_S \rvert=1$), a single spin flip cannot provide coupling between the near-degenerate hyperfine states of interest $\ket{a}$ and $\ket{b_1}$, which correspond to $M_S=1$ and $M_S=-1$, respectively (see above). Therefore, the coupling must involve an intermediate state $\ket{k}$, which is off-resonant by its Zeeman energy $\Delta_{ak}\simeq  1\mu_0\times 300$~G $\simeq$~400 MHz.  This further reduces the magnetic dipolar coupling between the open channels by the factor $(V^{mDD}/\Delta_{ak})^{-1}=7.7\times 10^6$  to much less than 1~mHz. 
We can thus rule out the magnetic dipolar interaction  as a source of the observed loss.

Spin flips in collisions of  $^3\Sigma$ molecules can also occur via coupling of the incident channels $\ket{a}$ and $\ket{b_1}$ to excited rotational states  \cite{Krems:04,Tscherbul:09}. This mechanism is similar to that of magnetic dipolar relaxation discussed above, with the excited rotational states ($N\ge 1$) playing the role of the intermediate Zeeman states.  A distinctive feature of this mechanism is that it is mediated by the anisotropy of the electrostatic interaction  between $^3\Sigma$ molecules (which couples the $N=0$ incident states to $N\ge 1$  closed-channel  states of the same $M_S$) and the spin-rotation and spin-spin interactions in the $N\ge 1$ manifolds, which couple states of different $M_S$.

Below we quantify this molecular spin relaxation mechanism by estimating the magnitude of the anisotropic coupling between  the degenerate open channels $\ket{a}$ and $\ket{b_1}$ due to the excited rotational states at $R=100 \text{ nm}$. We find that the strongest coupling due to the electric dipole-dipole interaction is only 0.05 kHz, and is therefore  too small to explain the observed loss rate.

\subsection*{Coupling matrix elements between degenerate open channels $\ket{aa}$ and $\ket{ab_1}$ due to rotationally excited states}

Here, we estimate  the matrix elements between the degenerate open channels $\ket{aa}$ and $\ket{ab_1}$ due to long-range interactions between NaLi(a$^3\Sigma^+$) molecules. The interactions are described by the multipole expansion \cite{StoneBook}

\begin{multline}
    \hat{V}(\mathbf{{R}}, \mathbf{r}_A, \mathbf{r}_B) = (4\pi)^{3/2}  \sum_{\lambda_A,\lambda_B,\lambda} V_{\lambda_A,\lambda_B,\lambda}(R,r_A,r_B) \\ \times A_{\lambda_A,\lambda_B,\lambda}(\hat{R},\hat{r}_A,\hat{r}_B)
\label{eq:multipole}
\end{multline}
where $A_{\lambda_A,\lambda_B,\lambda}(\hat{R},\hat{r}_A,\hat{r}_B)$ are the angular functions, $V_{\lambda_A,\lambda_B,\lambda}(R,r_A,r_B)$ are the radial expansion coefficients \cite{StoneBook}, $\hat{R}=\mathbf{{R}}/R$, and $\hat{r}_i=\mathbf{r}_i/r_i$. To leading order, the expansion \eqref{eq:multipole} contains the electric dipole-dipole, dipole-quadrupole, and quadrupole-quadrupole interactions. We assume that the internuclear distances of  NaLi molecules are fixed at their equilibrium values. The rigid rotor approximation is expected to be extremely accurate since the long-range NaLi-NaLi interactions at  $R=R_b$ (see below) are thousands of times smaller than the  spacing between the ground and the first excited vibrational states of NaLi ($\hbar\omega_{10}=40.2$ cm$^{-1}$ \cite{Gronowski:20}). Since we are interested in long-range physics outside the $p$-wave barrier, we will also neglect the spin dependence of the NaLi-NaLi interaction, which is significant only at very close range ($R\leq 10\, a_0$).

The radial expansion coefficients in Eq.~\eqref{eq:multipole} are expressed in terms of the multipole moments $Q_{\lambda_i,0}$ of the interacting molecules ($i=A, B$)

\begin{equation}
\begin{aligned}
    V_{\lambda_A,\lambda_B,\lambda}(R,r_A,r_B)  & =  \frac{Q_{\lambda_{A}0}Q_{\lambda_{B}0}}{R^{\lambda+1}}\delta_{\lambda,\lambda_{A}+\lambda_{B}} \\ &\cdot \frac{(-1)^{\lambda_A}}{[(2\lambda_{A}+1)(2\lambda_{B}+1)(2\lambda+1)]^{1/2}} \\ &\cdot
    \left[\frac{(2\lambda_{A}+2\lambda_{B}+1)!}{(2\lambda_{A})!(2\lambda_{B})!}\right]^{1/2}
\end{aligned}
\label{eq:coefficients}
\end{equation} 
The leading terms for two interacting neutral polar molecules such as NaLi are

\begin{equation}
\begin{split}
    V_{112}(R) & =- \frac{d^A d^B}{R^3} \sqrt{\frac{2}{3}} \text{ (dipole-dipole)},
   \\ V_{123}(R) & =- \frac{d^A Q^B_{20}}{R^4}  \text{ (dipole-quadrupole)},
   \\ V_{224}(R) & =\frac{Q^A_{20} Q^B_{20}}{R^5}\sqrt{\frac{14}{5}} \text{ (quadrupole-quadrupole)}.
\end{split}
\label{eq:coefficients leading}
\end{equation}
where $d^i$ and $Q^i_{20}$ are the electric dipole and quadrupole moments of the $i$-th molecule. Note that the long-range interaction \eqref{eq:multipole} is spin-independent, and hence can only couple the states of the same $M_S$, $M_{I_1}$, and $M_{I_2}$.  We use the accurate {\it ab initio} value of the molecule-frame electric dipole moment $d_\text{NaLi}=0.167$~D \cite{Gronowski:20}, and an approximate value of the electric quadrupole moment $\Theta_\text{NaLi}=10$ a.u. based on the calculated values for Na$_2$ and Li$_2$ from Ref.~\cite{Harrison:05}. Our estimates are not sensitive to the precise magnitude of $\Theta_\text{NaLi}$, since the dominant contribution at $R=R_b$ is given by the electric dipole-dipole interaction. 

\begin{figure}
    \centering
	\includegraphics[width = 83mm, keepaspectratio]{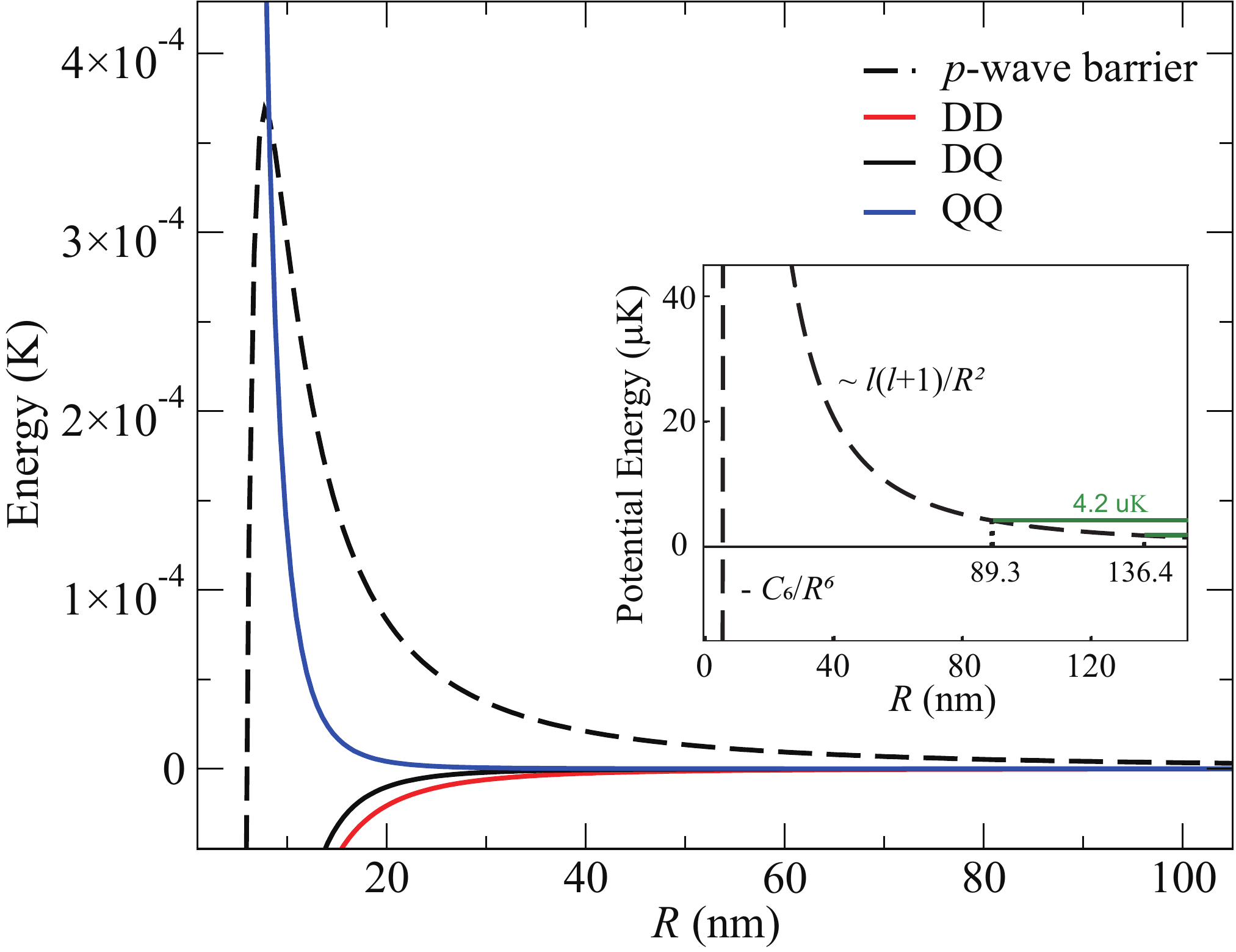}
	\caption{\textbf{ Radial dependence of the dipole-dipole, dipole-quadrupole, and quadrupole-quadrupole interactions of NaLi ($a^3\Sigma^+$) molecules.} The $p$-wave centrifugal barrier is also shown (dashed line). The upper and lower bounds on the experimental collision energies (4.2 $\mu$K and 1.8 $\mu$K) are marked by green horizontal lines. The outer turning points for these collision energies are $R_b=89.3$ and 136.4 nm, respectively.}
	\label{fig:LR interaction}
\end{figure}

\indent Figure \ref{fig:LR interaction} shows the radial dependence of the long range interactions between two NaLi molecules. While the dipole-dipole interaction dominates outside the $p$-wave barrier ($R \geq 100 \text{ nm}$) both the dipole-quadrupole and quadrupole-quadrupole interactions grow in magnitude as $R$ becomes shorter. At $R=100 \text{ nm}$, the magnitude of the electric dipole-dipole, dipole-quadrupole, quadrupole-quadrupole, and magnetic dipole-dipole interaction terms in Eq.~\eqref{eq:coefficients leading} are (in kHz): $V_{DD}= -3.446$, $V_{DQ}= -0.339$, $V_{QQ}= 0.0273$, and $V_{mDD}= 0.0519$. 

Equating the barrier energy plotted in Fig.~\ref{fig:LR interaction} and the endpoints of the experimental range of collision energies ($1.8{-}4.2$ $\mu$K), we obtain the corresponding range of distances of closest approach of two NaLi molecules in the $p$-wave channel as $R_b= 89.3{-}136.4$~nm.  For simplicity, we will use a value close to the middle of this interval, $R_b=100$~nm = 1890~$a_0$, to estimate the magnitude of all long-range coupling matrix elements.

Having parameterized the anisotropic long-range  interaction between two NaLi molecules  \eqref{eq:coefficients leading}, we now proceed to evaluate its matrix elements between the degenerate open channels  $\ket{a}$ and $\ket{b_1}$. The general matrix elements are given by:

\begin{equation}
    \bra{\gamma_A\gamma_B l m_l \eta} \hat{V}(\mathbf{R}, \mathbf{r}_A, \mathbf{r}_B) \ket{\gamma_{A}'\gamma_{B}' l' m_{l}' \eta'}
\label{eq:matrix elements}
\end{equation}
where $\bra{\gamma_A\gamma_B l m_l \eta}$ are the properly symmetrized basis states for two identical molecules ($\eta = -1$ for identical fermions), $\gamma_{A}$ and $\gamma_{B}$ refer to the internal hyperfine-Zeeman states of the molecules, $l$ is the orbital angular momentum for the collision, and $m_l$ is its projection on the space-fixed quantization axis defined by the external magnetic field \cite{Tscherbul:09}. The initial scattering state  of interest corresponds to $\gamma_A=\gamma_B=a$, $l=1$, $m_l=0$, and $\eta=-1$. 

The matrix elements are calculated by a straightforward extension of the procedure described in Ref.~\cite{Tscherbul:09} to include the hyperfine structure of both NaLi molecules (see the Supplemental Material of Ref.~\cite{Hermsmeier:21} for more details about the basis functions). 
Using a minimal basis including three lowest rotational states of each of the NaLi molecules ($N=0{-}2$) and two partial waves ($l=1,3$), leads to the total number of coupled channels $N_\text{ch}= 9908$ for the total angular momentum projection $M_\text{tot}=-7$.
 We note that this basis set is expected to produce converged results at $R=100$~nm, where the largest anisotropic long-range coupling $\lvert V_{DD} \rvert= 3.446$~kHz is much smaller than the spacing between the $N=0$ and $N=1$ rotational levels ($2B_e=8.4$~GHz).

\begin{figure}
    \centering
	\includegraphics[width = 83mm, keepaspectratio]{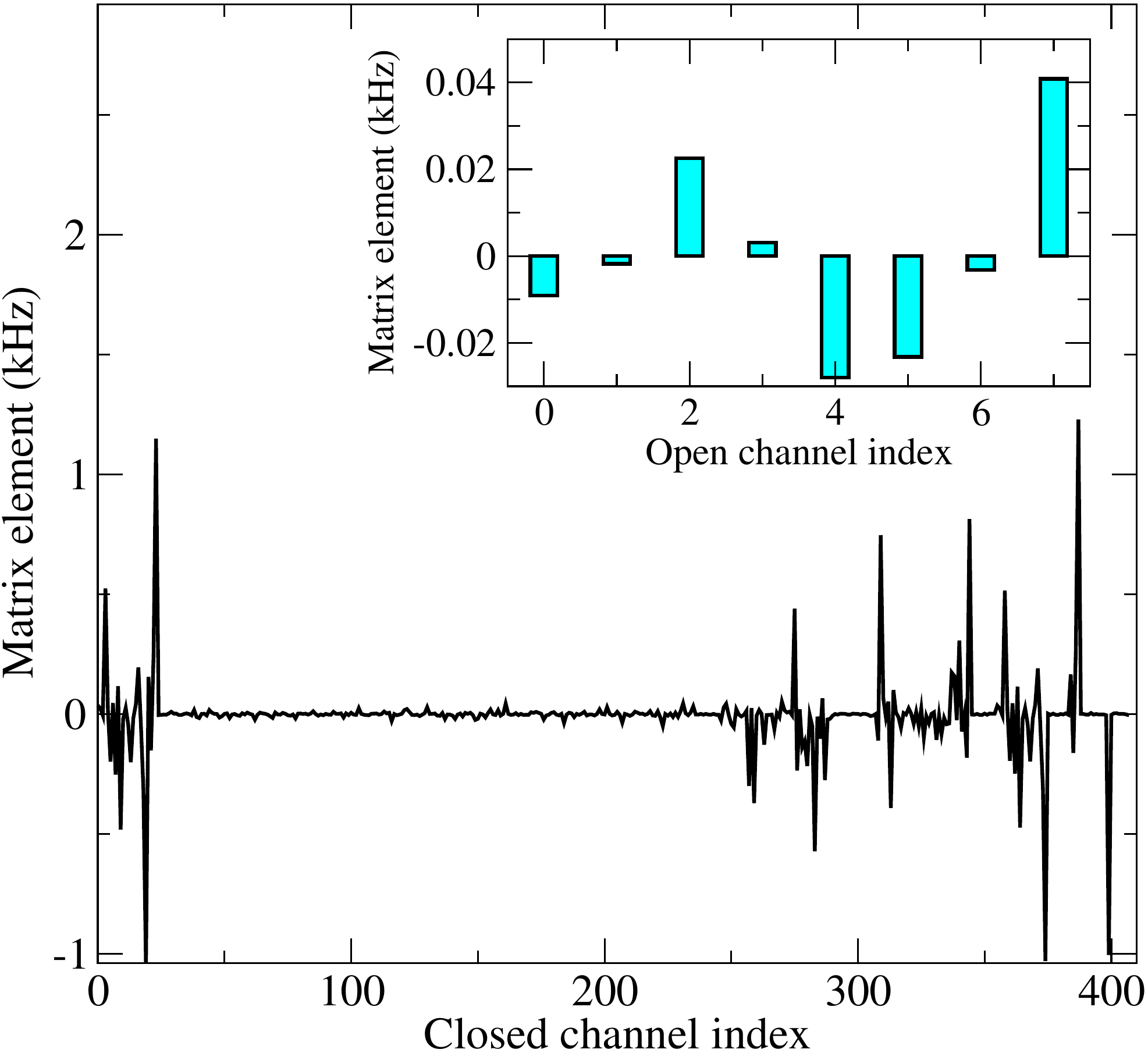}
	\caption{Matrix elements of the NaLi-NaLi interaction at $R = 100$~nm as a function of the channel index labeling the basis states $\ket{\gamma_{A} \gamma_{B} l m_l \eta}$. The initial channel is $\ket{aa, l = 1, m_l = 0}$ and the total angular momentum projection $M_\text{tot} = -7$. The channel index labels closed channels, in which one or both NaLi molecules are in their $N\ge 1$ excited rotational states. Only the matrix elements with the absolute magnitude exceeding 1 Hz are plotted. The magnetic field $B = 333$~G is tuned near the crossing between the $\ket{a}$ and $\ket{b_1}$ hyperfine-Zeeman levels. Inset: Histogram of direct  coupling matrix elements between the incident channel and lower-lying open channels, where both NaLi molecules are in the ground $N=0$ rotational states.}
	\label{fig:matrix element chart}
\end{figure}

\indent Figure \ref{fig:matrix element chart} shows the matrix elements between the incident channel $\ket{aa,l=1,m_l=0}$ at $R = 100 \text{ nm}$ and all final channels. The largest of these matrix elements $V_{01}\simeq$1.5~kHz is due to the long-range electric dipole-dipole coupling between the ground and excited ($N = 1$) rotational states. While these matrix elements do not directly couple the degenerate channels $\ket{a}$ and $\ket{b_1}$, they do contribute to indirect second-order couplings estimated below.
There are also direct couplings between the incident channel and lower-lying relaxation channels (with single-molecule hyperfine-Zeeman state lower in energy that the incident state $\ket{aa}$), which occur between the $l=1$ and $l=3$ partial waves. These couplings are mediated by the intramolecular spin-spin interaction, which couples the different $M_S$ components of the $N=0$ and $N=2$ rotational states \cite{Krems:04}. As shown in the inset of Fig.~\ref{fig:matrix element chart}, the largest of these direct couplings is about 0.04~kHz, which is too small to be responsible for the observed loss.  

It remains to consider the second-oder couplings between two degenerate channels $\ket{a}$ and $\ket{b_1}$ via rotationally excited states. These couplings are  suppressed by the factor $(V_{01}/\Delta_{01})^{-1}$, where $\Delta_{01}=2B_e$ is the energy of $N=1$ rotational states relative to the ground rotational state. Using the values $\Delta_{01}=8.4$~GHz and $V_{01}=1.5$~kHz, we obtain $(V_{01}/\Delta_{01})^{-1}=5.6\times 10^6$, and thus the second-order couplings are smaller than 1 mHz, and can be  neglected.

\subsection{Fabry–Pérot transmission}    
\indent The transmission of flux through a Fabry–Pérot cavity with two mirrors, M$_1$ and M$_2$ is exactly expressed by an Airy distribution in terms of the incoming light intensity, $I$, mirror reflection and transmission coefficients, $r_i$ and $t_i$, and single-pass phase shift $\phi$, as:

\begin{equation}    
   T_\text{trans}=I \frac{t_{1}^{2}t_{2}^{2}} {(1-r_{1}r_{2})^2 +  4r_{1}r_{2}\sin^2{\phi}}.
	\label{eq:Airy}
\end{equation} 
\noindent This Airy distribution is well approximated by a Lorentzian spectral line shape near a resonance ($\omega \oldsim \omega_{o}$) for highly reflective cavities ($r_{1}^{2}r_{2}^{2} \ll 1$) as:

\begin{equation}    
   T_\text{trans}=I \frac{\gamma_{1}\gamma_{2}} {(\omega_o-\omega)^2+[(\gamma_{1}+\gamma_{2})/2]^2}
	\label{eq:Lorenzian 2}
\end{equation} 
where $\gamma_{i} {=} -2\ln{r_i}/\tau_{RT}$ are the mirror coupling strengths, $\tau_{RT}$ is the round-trip time for a pulse travelling in the cavity, $\omega_{0}$ is the angular frequency of the cavity mode and $\omega$ is the angular frequency of the incoming light \cite{ismail2016fabry}.

Equation \ref{eq:Lorenzian 2} also describes the dissipation in a  harmonic oscillator driven by a friction force.  A harmonic oscillator with resonance $\omega_0$ and damping rate $\gamma_2$ driven at frequency $\omega$ via a frictional input coupling $\gamma_1$ is described by the differential equation:

\begin{equation}
\begin{split}
    \frac{d^2q(t)}{dt^2}+\gamma_{2}\frac{dq(t)}{dt}+\gamma_1 \frac{d}{dt}[q(t)-q_d(t)]+\omega_{0}^{2}q(t)=0. 
\end{split}
\label{eq:differential eq1}
\end{equation}
Here, we assume the drive $q_{d}(t) = q_{0}\sin\omega t$. By rearranging the terms, we obtain the standard equation of a driven harmonic oscillator with damping $\gamma=\gamma_1+\gamma_2$  

\begin{equation}
\begin{split}
    \frac{d^2q(t)}{dt^2} + \gamma\frac{dq(t)}{dt} + \omega_{0}^{2}q(t)= \gamma_1 \frac{d}{dt}q_d(t).
\end{split}
\label{eq:HO equation}
\end{equation}
The steady state solution of Eq.\ref{eq:HO equation} is

\begin{equation}
    q(t)= \frac{\gamma_1 q_{d}\omega}{(\omega^{2}-\omega_{0}^{2})^{2}+(\gamma\omega)^{2}}\cos(\omega t-\phi)
\label{eq:HO equation solution}
\end{equation}
where $\phi$ is defined by $\tan\phi=\gamma\omega/(\omega^{2}-\omega_{0}^{2})$.
The rate of energy dissipation due to $\gamma_2$ is $P_{out}=\gamma_2\dot{q}(t)^2$.
The average dissipation power

\begin{equation}
\begin{aligned}
    \left<P_{out}\right> & =\left<\gamma_{2} \dot{q}^2\right> \\
    & = \left< \gamma_2\frac{\gamma_1^2 q_{0}^2 \omega^4}{(\omega^2-\omega_{0}^2)^2+(\gamma\omega)^2}\sin^2(\omega t-\phi) \right> 
\end{aligned}
\label{eq:Dissipation power}
\end{equation}
is approximated to $\frac{1}{8}\frac{\gamma_1^2 \gamma_2 q_{0}^2\omega^2}{(\omega-\omega_{0})^2+(\gamma/2)^2}$ near resonance.
The ratio of the average dissipated power normalized by a quarter of the average nominal drive power (drive power at zero amplitude of the harmonic oscillator) is exactly given by Eq. \ref{eq:Lorenzian 2}.

\subsection*{Three-state $T$-matrix model of $p$-wave resonant scattering near degeneracies}

The main purpose of this section is to provide a microscopic derivation for Eq. \ref{eq:Loss rate} in the main text using an extended  nonperturbative $T$-matrix model  of $p$-wave resonant scattering \cite{chevy2005resonant}. The model includes a single $p$-wave bound state (or closed channel) $\ket{3}$ coupled to two open channels: the incident channel $\ket{1}$ and the outgoing inelastic channel $\ket{2}$ with threshold energies $E_{1}$ and $E_{2}$ ($E_{1} > E_{2}$) as illustrated in Fig. \ref{fig:cartoon}(a). The open channels are separated by the energy gap $\Delta > 0$ such that $E_{1} = E_{2} + \Delta$. The total energy of the two-molecule system before the collision is $E_{1}(k) = k^{2}/2\mu = k^{2}/m$, $\mu = m/2$ is the reduced mass of the two identical molecules of mass $m$, $k$ is the wavevector in the incident open channel $\ket{1}$, and we have set $E_{1} = 0$, so that $E_{2} = -\Delta$. The incident scattering state in channel $\ket{1}$ is a plane wave $\ket{\mathbf{k}}$ with energy $E_{1}(k) = k^{2}/m$ multiplied by the internal state vector of the colliding molecules $\ket{\alpha}$.

\begin{figure}[t!]
  \centering
	\includegraphics[width = 86mm, keepaspectratio]{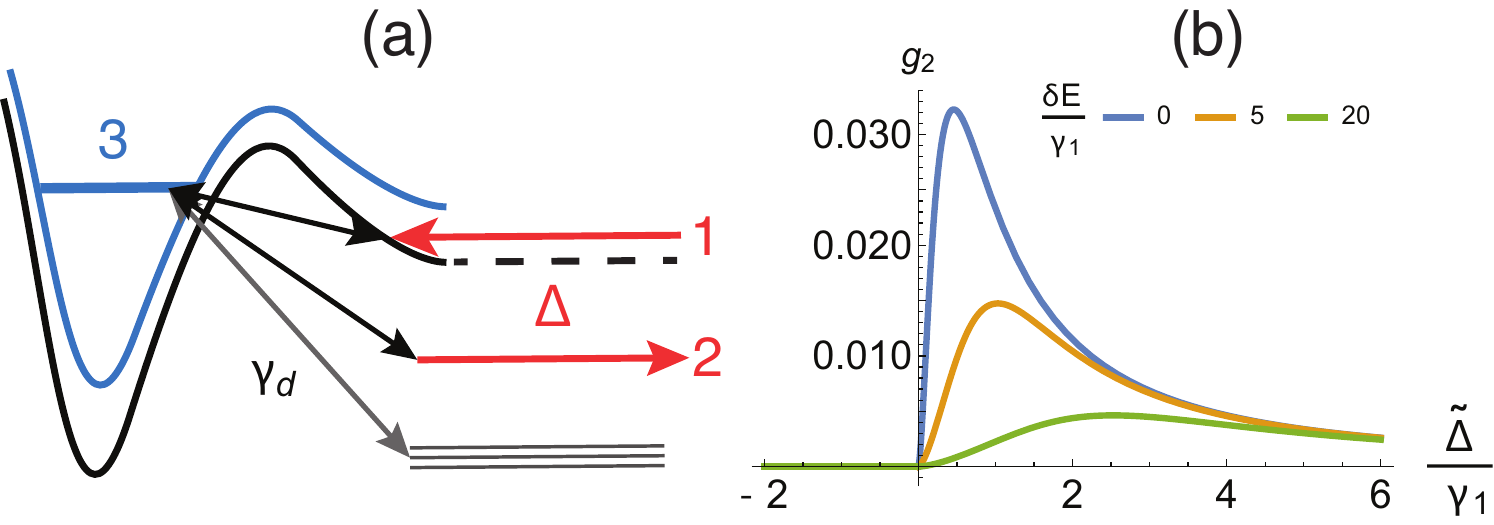}
	\caption{a. Schematic of the resonance model with two open channels and a p-wave bound state trapped behind a centrifugal barrier. b. {Inelastic rate $g_{2}(\tilde{\Delta})$ (in arbitrary units) plotted as a function of $\tilde{\Delta}/\gamma_{1}$ for the different values of detuning from resonance normalized by $\gamma_1$, $\delta E/ \gamma_1$. Note that for $\tilde{\Delta} < 0$ the channel $\ket{2}$ becomes closed and thus $g_2 (\tilde{\Delta}) = 0$.} }
	\label{fig:chevy model}
\end{figure}

By summing the diagrammatic expansion for the $T$-matrix, one obtains the following expression for the off-diagonal matrix elements between the open channels $\alpha$ and $\alpha'$ ($\alpha, \alpha' = 1, 2$) \cite{chevy2005resonant}

\begin{equation}    
   T_{\alpha,\alpha'}=\frac{C}{L^{3}} \frac{kk'F_{\alpha}(k)F_{\alpha'}(k')}{E-\delta-\Sigma_{1}-\Sigma_{2}},
	\label{eq:T matrix}
\end{equation} 
where $E=k^2/m$ is the collision energy in the incident channel $\ket{1}$ and $\delta > 0$ is the energy of the bare p-wave bound state $\ket{3}$, $L^3$ is the quantization volume, and $C=\cos\theta$, where $\theta$ is the angle between the incoming and outgoing wavevectors \cite{chevy2005resonant}. The functions $F_{\alpha}(k)$ quantify the coupling between the open and closed channels as a function of the wavevector $k$, and $\Sigma_{\alpha}(E) {=}\frac{4\pi}{3} (2\pi)^{-3}\int{q^{4}}dq \frac{\lvert F_{\alpha}(q) \rvert^{2}}{E-E
_{\alpha}(q)}$ are the open-closed channel couplings in the energy space. These couplings are crucial as they determine the resonance width. They can be evaluated by  regularizing  the diverging terms, and then setting $F(q) \rightarrow F(0)$, which results in the following expression

\begin{equation}    
\begin{split}
   \Sigma_{\alpha}(E) & =\lambda_{\alpha}\frac{-i}{12\pi} m(m[E-E_{\alpha}(0)])^{3/2} \\
   & - \delta_{0}^{(\alpha)} - \eta_{\alpha}[E-E_{\alpha}(0)]
\end{split}
\label{eq:coupling}
\end{equation}
where $\lambda_{\alpha}=\lvert F_{\alpha}(0) \rvert^{2}$ are the $q \rightarrow 0$ limits of the open-closed channel couplings, and the integrals

\begin{equation}    
   \delta_{0}^{(\alpha)}= (6\pi^{2})^{-1}\int \lvert F_{\alpha}(q) \rvert^{2}mq^{2}dq
   \label{eq:coupling 1}
\end{equation} 
\begin{equation}    
   \eta_{\alpha}= (6\pi^{2})^{-1}\int \lvert F_{\alpha}(q) \rvert^{2}mdq
   \label{eq:coupling 2}
\end{equation} 

\noindent depend on the exact form of the coupling matrix elements between the open and closed channels. Note that (i) $E_{1}(0) = 0$ and $E_{2}(0) = -\Delta$ by definition, and (ii) the first term on the right-hand side of Eq. (\ref{eq:coupling}) is purely imaginary (since we assume $\Delta > 0$) and thus gives rise to resonance width 

\begin{equation}
\begin{split}
   \gamma(E,\Delta) {=} & (\gamma_{1} + \gamma_{2}) + \gamma_{d}  \\
   = & -2 \mathfrak{Im}(\Sigma_{1}+\Sigma_{2}) + \gamma_{d}  \\
   = & \frac{m^{5/2}}{6\pi} (\lambda_{1}E^{3/2} + \lambda_{2}(\Delta+E)^{3/2}) + \gamma_{d}
    \end{split}
\label{eq:width}
\end{equation} 
where we have introduced the intrinsic width $\gamma_{d}$ of the $p$-wave bound state due to the coupling to lower-lying inelastic channels other then $\ket{1}$ and $\ket{2}$ [see Figure \ref{fig:chevy model}(a)].  While Ref.~\cite{chevy2005resonant} assumes that $\Delta$ is much larger than all the other energy scales in the problem, we do not make such an  assumption here.  Indeed, in our model, the new Feshbach resonance occurs when $\Delta \rightarrow 0$.

Defining the resonance shift $\delta_{0} = \delta_{0}^{1} + \delta_{0}^{2}$ and neglecting the dimensionless terms $\eta_{\alpha}$, which are expected to be very small compared to unity \cite{chevy2005resonant}, we obtain from Eq. \eqref{eq:T matrix}
 
\begin{equation}    
   T_{\alpha,\alpha'}=\frac{C}{L^{3}} \frac{kk'F_{\alpha}(0)F_{\alpha'}(0)}{E-(\delta-\delta_{0}) + i\gamma(E,\Delta)/2}.
	\label{eq:T matrix 2}
\end{equation} 
Here, we have also assumed that the bound-continuum coupling matrix elements $F(k)$ are well approximated by their zero-$k$ values $F_{\alpha}(0)$, which is a good approximation in the limit $k \rightarrow 0$ (note, however, that this approximation starts to break down as $kR_{3}$ approaches 1, where $R_{3}$ is the ``size'' of the $p$-wave bound state, as shown below). The final wavevector in Eq. (\ref{eq:T matrix 2}) $k' =  \sqrt{m(\Delta + E)}$.
We are interested in the two-body inelastic rate for the transition $\ket{1} \rightarrow \ket{2}$ at fixed collision energy, which is given by (up to a constant scaling factor)

\begin{widetext}
\begin{equation}
   g_{2}(E,\Delta)= |T_{1, 2}|^{2}\rho_{2}(k') 
   \simeq \frac{k^{2} m(\Delta + E)^{3/2}\lambda_{1}\lambda_{2}}{[E-(\delta-\delta_0)]^2 +  [\frac{m^{5/2}}{6\pi}\{\lambda_{1}E^{3/2} + \lambda_{2}(\Delta+E)^{3/2}\}+\gamma_d]^{2}/4}
\label{eq:inelastic rate}
\end{equation}
\end{widetext}
where $\rho_{2}(k') = (m/2)^{3/2}\sqrt{2(\Delta+E)}/2\pi^2$ is the density of states in the final channel $\ket{2}$ \cite{FriedrichBook}.   
Eq. (2) of the main text is identical to {\it Eq.~(\ref{eq:inelastic rate}) up to a constant overall scaling factor and with the intrinsic decay width of state $\ket{3}$ $\gamma_d=0$.  This provides a microscopic justification for the Fabry-P\'erot model}. In particular, the Fabry-P\'erot decay rates may be expressed as $\gamma_1=\frac{m^{5/2}}{6\pi}\lambda_{1}E^{3/2}$ and $\gamma_2=\frac{m^{5/2}}{6\pi}\lambda_{2}(\Delta+E)^{3/2}$, providing insight into their collision energy and $\Delta$ dependence.

\indent We now discuss the main features of the expression for the two-body inelastic rate \eqref{eq:inelastic rate}. 
To this end, consider the expression for the resonance width $\gamma(E,\Delta)$ in the denominator of Eq.~(\ref{eq:inelastic rate}) given by Eq.(\ref{eq:width}).  In addition to the intrinsic width  $\gamma_d$, the width contains contribution from (i) the coupling between the incident open channel $\ket{1}$ with the $p$-wave bound state $\gamma_1 \propto E^{3/2}\propto k^3$, and (ii) the coupling between open channels $\ket{1}$ and $\ket{2}$ through the bound state $\gamma_2 \propto (E+\Delta)^{3/2}\propto (k')^3$.

\indent
Equation~\eqref{eq:inelastic rate} shows that the  inelastic rate away from the resonance (when $E - (\delta-\delta_0)> \gamma_1 + \gamma_2$) or for $\gamma_d >  \gamma_1 + \gamma_2$ exhibits the standard  $p$-wave scaling  $g_2\propto k^2 (k')^3$ \cite{chevy2005resonant} as observed experimentally. The normal $p$-wave threshold scaling will be modified if the denominator of Eq.~\eqref{eq:inelastic rate} becomes energy-dependent, which requires the detuning $E -(\delta-\delta_0)$ and $\gamma_d$ to be small compared to $\gamma_1+\gamma_2$. Under these (rather unlikely) conditions, the scaling changes to $g_{2}(E) \propto k^2(k')^3/[\lambda_1 k^{3} + \lambda_2 (k')^{3}]$.

To further explore the properties of the inelastic rate in Eq.~\eqref{eq:inelastic rate},  it is convenient to introduce the parameter $\tilde{\Delta}=\Delta+E$, such that  the  translational energy in the outgoing channel $\ket{2}$ vanishes at $\tilde{\Delta}=0$ ($k'=0$), and inelastic scattering becomes energetically forbidden at $\tilde{\Delta}<0$. With this definition, we obtain from Eq.~\eqref{eq:inelastic rate} the inelastic rate as a function of $\tilde{\Delta}$ 

\begin{equation}
 g_{2}(\tilde{\Delta}) \simeq
    \frac{(\tilde{\Delta})^{3/2}}{\delta E^2 +  [\gamma_1 + c\tilde{\Delta}^{3/2}]^2/4},
\label{eq:inelastic rate fixed E}
\end{equation}
where $\delta E =E-(\delta-\delta_0)$ is the detuning from resonance, $c=\frac{m^{5/2}}{6\pi}\lambda_2$, and we have omitted the factors proportional to $k$, $m$ and $\lambda_i$ in the numerator since we are interested in the inelastic rate at a fixed collision energy. 

Note that our ability to vary $\delta E$ by tuning an external magnetic field may be limited, since the magnetic moments  of the closed-channel and open-channel $p$-wave states may be very close (as in the case of a single-channel $p$-wave shape resonance). We therefore  do not  assume the resonance condition, and keep the term  $\delta E^2$  in Eq.~\eqref{eq:inelastic rate fixed E}.

Figure \ref{fig:chevy model}(b) shows that the inelastic rate $g_2(\tilde{\Delta}$) displays a pronounced resonance structure as a function of $\tilde{\Delta}$. The remarkable increase of the inelastic rate with narrowing the energy gap between the open channels is a consequence of the reduction of the total resonance width (\ref{eq:width}) in the limit $\Delta \rightarrow 0$  where $\gamma_2\to 0$ (note that we also require that $\gamma_d\ll \gamma_1$). This reduction enhances the peak rate of inelastic decay of the $p$-wave bound state into channel $\ket{2}$ above the universal limit.

The resonance profiles shown in Fig.~\ref{fig:chevy model}(b) are the sharpest at zero energy detuning ($\delta E=0$), where the denominator of Eq.~\eqref{eq:inelastic rate fixed E} is most sensitive to  $\tilde{\Delta}$. As expected, the resonance becomes more and more suppressed as one moves away from resonance due to the growing background contribution from the $\delta E^2$ term in Eq.~\eqref{eq:inelastic rate fixed E}.
In principle, the background contribution can arise not only from a finite $\delta E$, but also from other mechanisms, such the intrinsic decay of the $p$-wave bound states to deeply bound channels (parametrized by  $\gamma_d$), which may contribute to the overall decay rate in Eq.~\eqref{eq:width}.

In the experiment, the temperature and the observed width of the Feshbach resonance are similar. The strong observed dependence on $\Delta$ thus suggests that the  alternative decay mechanisms are slow compared to the dominant decay channels, {\it i.e.}, $\gamma_d \lesssim \gamma_i$ ($i=1,2$).
  
\subsection*{Derivation of $F_{\alpha}(k)$ matrix elements}

\indent The $p$-wave bound state in channel $\ket{3}$ is described by the wavefunction $\langle\mathbf{R} \vert \psi_{3}\rangle = \psi_{3}(\mathbf{R})$ of the intermolecular separation vector $\mathbf{R}$. For a single $p$-wave bound state, the radial and angular variables separate to give $\psi_{3}(\mathbf{R})= g_{1}(R)Y_{1m_{\mathbf{u}}}(\hat{R})$, where $m_{\mathbf{u}}$ is the projection of the bound state's angular momentum $l=1$ on a quantization axis $\mathbf{u}$. As in Ref.~\cite{chevy2005resonant}, we assume that the bound state is coupled to both of the open channels $\ket{\mathbf{k}, \alpha}$ ($\alpha = 1,2$) via the coupling matrix elements $\bra{\mathbf{k}, \alpha} V \ket{3 m_{k}}$, where $\lvert \mathbf{R} \vert \mathbf{k}\alpha \rangle =  e^{ikz}\ket{\alpha}$ are the scattering states in the open channels $\alpha$. Expanding the incoming plane wave in spherical waves, $e^{ikz}=\sum_{l} i^{l}\sqrt{4\pi(2l+1)} j_{l}(kR) Y_{l0}(\theta, \phi)$, where $j_l(kR)$ is a spherical Bessel function, we observe that only the $p$-wave  ($l=1$) components of the incident and outgoing waves couple to the $p$-wave bound state due to the orthogonality of the spherical harmonics (and the assumption of isotropic bound-continuum coupling). 

\indent For practical purposes it is convenient to define the bound-continuum couplings in $k$-space \cite{chevy2005resonant}

\begin{equation}
    F_{\alpha}(k)= i\sqrt{12}\bra{\alpha}\hat{V}\ket{3} \frac{1}{k} \int_{0}^{\infty} g_{1}^{*}(R)j_{1}(kR)R^{2}dR
\label{eq:F}
\end{equation}
where $\bra{\alpha}\hat{V}\ket{3}$ is a spin matrix element. The advantage of the $F_{\alpha}(k)$ matrix elements is that they have a well-defined $k \rightarrow 0$ limit. Expanding $j_{1}(kR)\simeq kR/3 + O((kR)^{5})$, we get

\begin{equation}
    F_{\alpha}(0)= i\sqrt{12}\bra{\alpha}\hat{V}\ket{3} \frac{1}{k}\int_{0}^{\infty}g_{1}^{*}(R)R^{3}dR
\label{eq:F0}
\end{equation}
The approximation $F(k) \simeq F(0)$ is used in the previous section and in Ref. \cite{chevy2005resonant} to simplify the expression for the $T$-matrix elements near threshold. This approximation is valid as long as $kR_{3}\leq 0.1$, where $ R_{3}$ is the characteristic size of the $p$-wave bound state [$g_{1}(R) \simeq R^{2}e^{-R/R_{3}}$]. We find that $F(k) \simeq F(0)$ is a good approximation for the incident collision channel ($kR_{3} \simeq 0.1$ for  $R_3 = 100a_{0}$ and   $E = 10$ kHz). This is no longer the case when  the open-closed splitting becomes large compared to the collision energy ($\Delta/E \gg 30$) or the $p$-wave bound state becomes extremely delocalized, in which case the exact $k$-dependent matrix element $F(k)$ should be used.

\end{document}